\documentclass[12pt,a4paper]{article}

\usepackage{jheppub}

\usepackage[latin1]{inputenc}
\usepackage{amsmath}

\usepackage{amsfonts}
\usepackage{physics}
\usepackage{amssymb}
\usepackage{faktor}
\usepackage{graphicx}
\usepackage{tikz}
\usepackage{amsthm}
\numberwithin{equation}{section}
\usepackage[english]{babel}
\usepackage{mathtools}
\usepackage{nicefrac}
\usepackage{lscape}
\usepackage{textgreek}
\usepackage{parskip}
\usepackage{enumerate}

\usepackage{tcolorbox}

\usepackage{todonotes}

\usepackage{dsfont}

\usepackage{url}

\usepackage{subcaption}
\usepackage{cleveref}

\hypersetup{
	pdftitle={Semiclassical Lyapunov exponent},
	pdfauthor={Fabian Haneder}}
\usepackage{upgreek}
\usepackage{array}
\usepackage{bm}

\author[a]{Fabian Haneder,}
\author[a]{Gerrit Caspari,}
\author[a]{Juan Diego Urbina,}
\author[a]{and Klaus Richter}
\affiliation[a]{Institut f\"ur Theoretische Physik, 
Universit\"at Regensburg, \\
Universit\"atsstr. 31, D-93053 Regensburg, Germany}

\emailAdd{fabian.haneder@ur.de}
\emailAdd{gerrit.caspsari@ur.de}
\emailAdd{juan-diego.urbina@ur.de}
\emailAdd{klaus.richter@ur.de}
\title{ 
The relation between classical and quantum Lyapunov exponent and the bound on chaos in classically chaotic quantum systems
}

\abstract{
    Out-of-Time-Ordered Commutators (OTOCs), representing a key diagnostic for scrambling as a facet of short-time quantum chaos,  have attracted wide-ranging interest, from many-body physics to quantum gravity. By means of a suitable form of the Wigner-Moyal expansion, and invoking ensemble equivalence in statistical physics, we provide a consistent approach to the growth rate of the OTOC for many-body systems with chaotic classical limit where both the classical Lyapunov exponent and the quantum nature of the density of states enter. Applying this construction to quantized high-dimensional hyperbolic motion, \textit{i.e.,} a quantum chaotic system that exhibits gravity-like correlation functions in the late-time regime, we compute the OTOC growth rate $\Lambda$ as a function of the number of degrees of freedom, $f$, and inverse temperature, $\beta$.
    We show that the scaled growth rate, $\Lambda/f$, can be described by a universal function of $f \beta$ and displays a cross-over from classical to quantum behavior as we increase $f$ and/or lower the temperature. In the deep quantum regime of infinite $f$, we find maximally fast scrambling in the sense of the Maldacena-Shenker-Stanford bound on chaos. This elucidates the non-perturbative mechanism underlying the saturation of the bound via quantum contributions to the mean density of states, and it provides further support for this dynamical system as a dual to two-dimensional quantum gravity. In this way, we present first evidence of maximally fast scrambling in a quantum chaotic system with a well-defined classical Hamiltonian limit, without invoking any external mechanism such as (disorder) averaging.
}
\makeindex

\begin{document}
\maketitle
\section{Introduction}
In recent years, the study of quantum chaos \cite{Haake2010a}, and in particular, of scrambling as a manifestation thereof in systems with a large number of degrees of freedom, has drawn interest from varied, a priori seemingly disparate fields of physics. Scrambling in this context stems from the spreading of initially localized correlations across many or all available degrees of freedom of a system -- strongly scrambling systems are in a sense the quantum analogs of classical systems with mixing Hamiltonian flows\footnote{With the important caveat that quantum scrambling originates from a unitary time evolution and is therefore reversible, while classical mixing is not.}. For a recent introduction to just some of the literature about scrambling, see \cite{touil_information_2024}. One important application is the study of black holes: in a by now famous thought experiment due to Hayden and Preskill \cite{hayden_black_2007}, quantum information falling into a black hole is scrambled across the horizon, and then quite rapidly reemerges as Hawking radiation; the time scale for this process is given by the so-called \emph{scrambling time} $t^*$, roughly the time scale necessary for initial state information to propagate
through the entire system. It was conjectured by Sekino and Susskind \cite{Sekino2008} that black holes are the fastest scramblers occurring in nature\footnote{So fast indeed that it might pose a problem for the no-cloning theorem \cite{wootters_single_1982,dieks_communication_1982}.}, with a scrambling time of the order of 
\begin{equation}
    t^*=C\beta\log(S),\label{eq:scrambling-time}
\end{equation}
where $\beta$ is the inverse Hawking temperature, $C$ is some numerical constant and $S$ the entropy of the system. This form was later shown to be universal with $C=\hbar/2\pi$ in \cite{Shenker2014a}. 

A more recent facet of the study of scrambling and the Hayden-Preskill protocol in the context of black holes concerns teleportation of quantum information \cite{yoshida_efficient_2017}, in particular across traversable wormholes \cite{gao_traversable_2021,jafferis_traversable_2022,kobrin_experiments_2025}. There, the Sachdev-Ye-Kitaev (SYK) model \cite{sachdev_gapless_1993,kitaev_alexei_nodate,noauthor_alexei_nodate}, the low-energy dual of two-dimensional Jackiw-Teitelboim (JT) gravity \cite{Jackiw1985,Teitelboim1983}, is exploited to study signatures of the quantum chaotic nature of the wormhole setup. Correlation functions of Hermitian operators serve as key diagnostics for this purpose, and a particularly important one is the out-of-time ordered commutator (OTOC).

The OTOC is a quantity of great interest in the study of quantum chaos, going back to the work of Larkin and Ovchinnikov \cite{larkin_quasiclassical_1969}. For a given quantum system with a Hamiltonian $H$, consider Hermitian operators $V,W$, and denote time-evolved operators by $W_t=e^{\frac{i}{\hbar}Ht}We^{-\frac{i}{\hbar}Ht}$.  Then, the OTOC reads
\begin{equation}
    \mathcal{C}(t)=-\ev{[W_t,V]^2},\label{eq:otoc}
\end{equation}
where $[\cdot,\cdot]$ is the commutator and $\ev{\cdot}=\tr(\frac{e^{-\beta H}}{\tr(e^{-\beta H})}\cdot)$ is the expectation value in the thermal state at temperature $T=1/k_B\beta$. In some of the literature, OTOC refers to the closely related out-of-time ordered \emph{correlator} $F(t)=\ev{W_tVW_tV}$, but we will mean only the commutator in this work.

In many quantum chaotic systems the OTOC is characterized by an initial exponential growth for times shorter than the scrambling time $t^*$ of the system
(although there are exceptions, see \cite{garcia-mata_out--time-order_2023} and references therein).
The rate of this exponential growth, $\Lambda$, defines the system's temperature-dependent quantum Lyapunov exponent,
\begin{equation}
\label{eq:quantumOTOC}
\mathcal{C}(t)\sim e^{\Lambda t}.
\end{equation}
We should note here that there are differing conventions in the literature on whether to call the growth exponent of $\mathcal{C}(t)$ $\Lambda$ or $2\Lambda$. The latter seems to be more common in the quantum chaos community \cite{garcia-mata_out--time-order_2023}, while the former is more prevalent in quantum gravity contexts \cite{Maldacena2016,turiaci_inelastic_2019,jahnke_chaos_2019}, and it is also the convention we will be adopting hereinafter.

\begin{figure}
    \centering
    \begin{subfigure}[t]{0.48\textwidth}
        \centering
        \includegraphics[width=\linewidth]{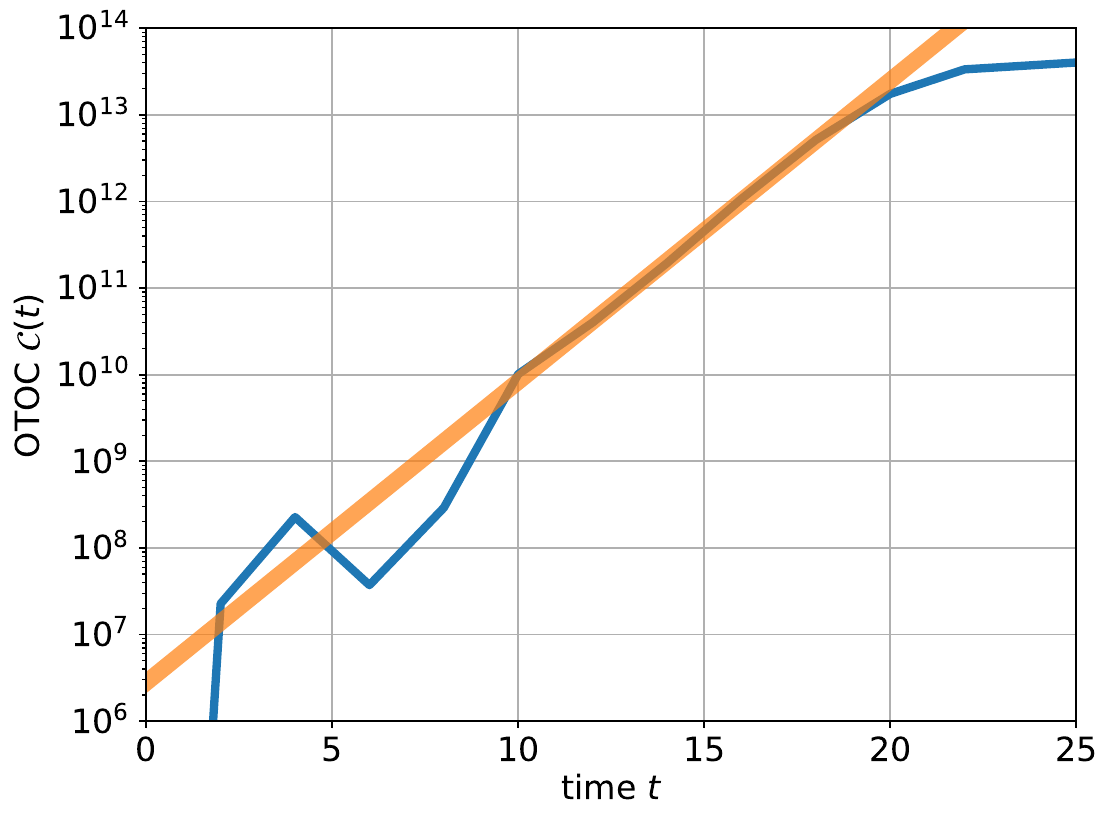}
        \label{fig:otoc-early}
    \end{subfigure}
    ~
    \begin{subfigure}[t]{0.48\textwidth}
        \includegraphics[width=\linewidth]{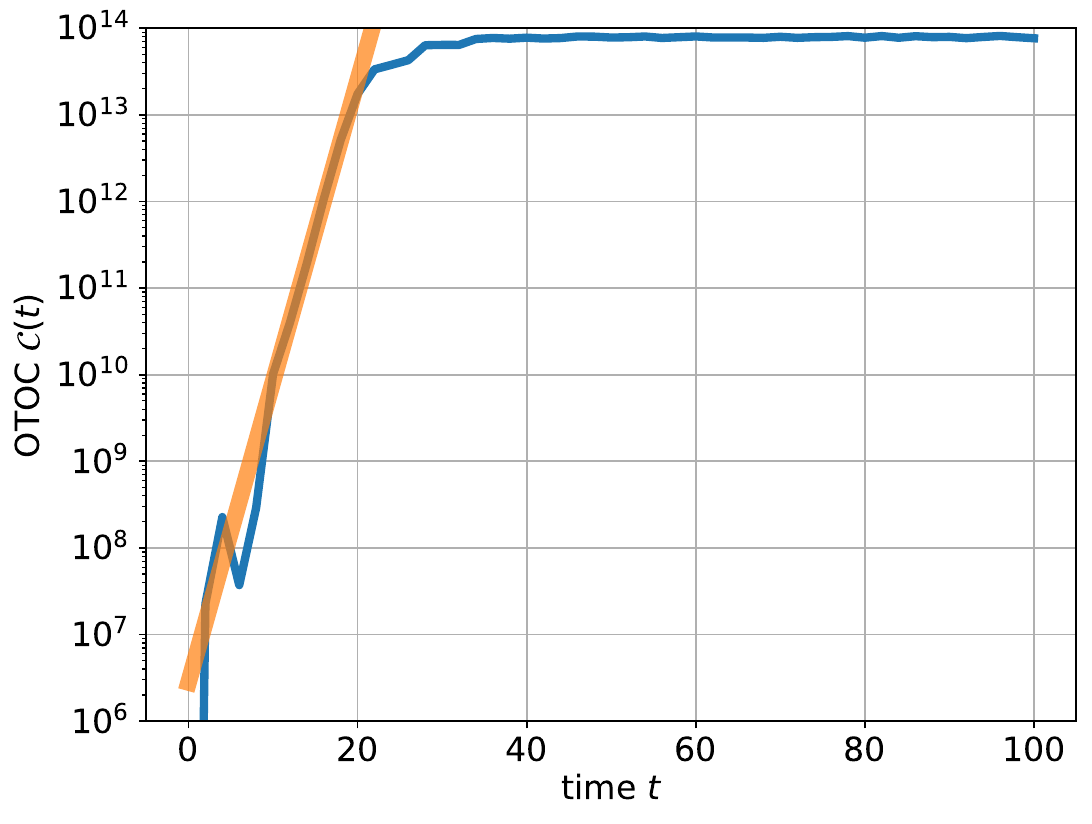}
        \label{fig:otoc-late}
    \end{subfigure}
    \caption{The OTOC $\mathcal{C}(t)=\ev{\abs{[\hat{n}_1(t),\hat{n}]}^2}_\psi$ in a strongly kicked Bose-Hubbard system with $N=10000$ particles and $L=2$ sites,
    where $\hat{n}_1$ is the particle number operator on the first site and $\psi$ is a suitably localized coherent state at energy $E$.  Left: early times. The nearly linear slope indicates the initial exponential growth with a constant rate corresponding to twice the classical Lyapunov exponent. Right: The OTOC saturates at late times. Blue solid line: numerical data. Orange opaque line: linear fit. Taken with permission from M. Steinhuber \cite{steinhuber}.}
    \label{fig:otoc}
\end{figure}

The reason for the nomenclature becomes obvious by picking for the operators e.g. a canonical pair $X,P$ and corresponding classical phase-space degrees of freedom $x,p$. Provided the system has a classical limit, heuristic application of canonical quantization demands that
\begin{equation}
   \ev{[X_{t},P]}_\psi\sim i\hbar\{x_t,p\},
\end{equation}
where the expectation value is defined with respect to a suitably localized state $\psi$ near a fixed energy $E$, and $x_t$ the initial coordinate $x$ evolved for a time $t$. Since for systems displaying chaotic dynamics in their classical limit,
\begin{equation}
\label{eq:classOTOC}
    \{x_t,p\}=\pdv{x_t}{x}\sim e^{\lambda t},
\end{equation}
where $\lambda(E)$ is the Lyapunov exponent of the classical counterpart, the corresponding OTOC, in this \textit{microcanonical} framework, initially increases exponentially with a rate $2\lambda$, as illustrated in \cref{fig:otoc}. More generally, 
 in this microcanonical situation, the exponential OTOC growth rate has been shown to agree with twice the classical $\lambda(E)$  \cite{jalabert_semiclassical_2018,rammensee_many-body_2018,garcia-mata_out--time-order_2023}. 
(For times longer than $t^*$, the OTOC saturates, as evident e.g. in \cref{fig:otoc} and predicted in \cite{rammensee_many-body_2018}.)

Also, \cref{eq:quantumOTOC,eq:classOTOC} heuristically suggest $\Lambda=2\lambda$. 
The key fact that the quantum Lyapunov exponent in eq.~(\ref{eq:quantumOTOC}) is defined by a canonical initial state (and cannot be directly related to $\lambda(E)$) has been a source of confusion particularly within the study of quantized chaotic systems by means of semiclassical methods, where one often works at fixed energy $E$ and hence with microcanonical quantities. Two notable exceptions are the work of Jalabert \emph{et al.} for systems with few degrees of freedom \cite{jalabert_semiclassical_2018}, and Hashimoto \emph{et al.} for general scaling systems \cite{hashimoto_bound_2022}. In the former, the temperature-dependent exponent is calculated by applying the statistical definition of mean energy in the canonical ensemble to the microcanonical classical exponent $\lambda(E)$. However, as we will show later, the use of the \emph{classical} Thomas-Fermi approximation for the microcanonical level density automatically renders the result classical, and the quantum features of $\Lambda(\beta)$ totally absent. In fact our analysis shows that the very definition of $\Lambda(\beta)$ in systems with few degrees of freedom (such as the ones considered in \cite{jalabert_semiclassical_2018} and \cite{pappalardi_low_2022,pappalardi_quantum_2023}) is ambiguous because the emergence of the exponential form of the OTOC will strongly rely on the standard assumptions behind the ensemble equivalence in statistical mechanics. Any discussion about the bound on chaos, that emerges only for systems with large number of degrees of freedom, is therefore out of reach unless further quantum effects and/or  very non-generic features of specific systems are included \cite{pappalardi_low_2022}. The analysis presented in \cite{hashimoto_bound_2022} is closer in spirit to our approach here, as it applies to systems with many degrees of freedom, but suffers from the lack of precise analytical results for $\lambda(E)$ and  absence of quantum effects in the many-body level density, rendering $\Lambda(\beta)$ out of reach again.

In this work, we propose a consistent semiclassical theory of \textit{canonical quantum} Lyapunov exponents in quantum systems with large number of degrees of freedom that admit a proper classical limit. This general approach will critically depend on two ingredients, namely the exact energy dependence of the \textit{classical microcanonical} Lyapunov exponent, and the full quantum mechanical mean level density. These two aspects of the classical and quantum description can be rigorously and explicitly  considered for the dynamics of a particle sliding on a high-dimensional hyperbolic manifold, a particular system that displays all the genuine properties of classical chaos while admitting an exact semiclassical quantization, as we will explain in detail later. 
This system has been shown in \cite{haneder_beyond_2025} to exhibit an emergent quantum-gravitational description in the sense that, in the limit of large configuration space dimension, \textit{i.e.} large number of degrees of freedom, its spectral correlation functions are identical to that describing a consistent theory of quantum gravity in low dimensions: Jackiw-Teitelboim (JT) gravity. We will show that this system is a fast scrambler and, in the correct limit, indeed a maximally fast scrambler in the sense of saturating the Maldacena-Shenker-Stanford (MSS) bound on chaos for the growth rate \cite{Maldacena2016},
as expected for a system dual to JT gravity. 

Since this system has a well-defined classical limit, we will follow a natural approach for the calculation of the quantum canonical Lyapunov exponent from a semiclassical perspective, where quantum properties of the system are appropriately expressed through its classical phase space structures. Due to the great degree of analytical control, afforded e.g.\ by the semiclassical description via the Selberg trace formula \cite{bytsenko_quantum_1996}, as well as the exactly known density of states and microcanonical Lyapunov exponent, we can not only rigorously show the saturation of the MSS bound, but also that this saturation is a quantum effect. It stems from quantum corrections to the leading power law behavior of the density of states. 
Moreover, for the system at hand, we provide  explicit results for the quantum Lyapunov exponent as a function of temperature and number of degrees of freedom, beyond the limiting cases.
We will further take initial steps towards evaluating the first subleading $\hbar^2$ correction to the OTOC, which can be systematically determined by the mechanism of Wigner-Moyal phase space quantization, and which has to exhibit exponential behavior with a different, likewise bounded growth exponent by the arguments of \cite{kundu_subleading_2022}. 

The rest of the paper is organized as follows: in \cref{sec:otoc}, we quickly recapitulate the MSS bound. In \cref{sec:wigner-moyal}, we introduce the formalism of Wigner-Moyal quantization, and particularly a way to obtain $\hbar$ expansions for Heisenberg operators developed in \cite{osborn_moyal_1995}. In \cref{sec:selberg}, we introduce the aforementioned system whose Lyapunov exponent we will compute in \cref{sec:leading,sec:corrections}. 
\section{The Maldacena-Shenker-Stanford bound}\label{sec:otoc}
A particularly interesting facet of the study of OTOCs is the appearance of an analytical bound on its decay rate \cite{Maldacena2016}. Namely, it can be shown that $f(t)=F(t)/F_d$, with $F_d=\ev{VV}\ev{WW}$ the (constant) factorized value of the out-of-time ordered correlator, satisfies
\begin{enumerate}
    \item $f(t+i\tau)$ is analytic in the half strip $0<t,~-\frac{\beta}{4}\leq\tau\leq\frac{\beta}{4}$ and real for $\tau=0$.
    \item $\abs{f(t+i\tau)}\leq1$ in the entire half-strip.
\end{enumerate}
For such a function, then, 
\begin{equation}
    \frac{1}{1-f}\abs{\frac{df}{dt}}\leq\frac{2\pi}{\hbar\beta},
\end{equation}
which implies, upon \textit{assuming} the form 
\begin{equation}
F(t)=F_d-\epsilon e^{\Lambda t},\label{eq:corr-leading}
\end{equation}
that the decay rate $\Lambda$ must obey
\begin{equation}
    \Lambda\leq\frac{2\pi}{\hbar\beta},\label{eq:mss-bound}
\end{equation}
which is the famous Maldacena-Shenker-Stanford (MSS) bound\footnote{We emphasize for clarity that the decay rate of $F(t)$, and thereby also the decay rate of the OTOC $\mathcal{C}(t)$, is bounded by $\frac{2\pi}{\hbar\beta}$, irrespective of whether we call this decay rate $\Lambda$ or $2\Lambda$.}. The proof of \eqref{eq:mss-bound} applies under fairly general assumptions, namely that the correlation functions factorize in a long enough time regime, and that there is a strong hierarchy between the scrambling time $t^*$ and the (shorter) \emph{dissipation time} $t_d$, which is roughly the decay time of two-point correlators like $\ev{VV_t}$. This hierarchy is expected particularly in systems with many degrees of freedom and Hamiltonians built from finite products of simple operators. This is the case in many typical many-body systems, and also in large-dimensional few- or one-body systems, which will be relevant in this work.

It is important to realize that the bound applies under the assumption that there is a physical mechanism leading to the exponential growth of the second term in \cref{eq:corr-leading}. Therefore, verifying its validity starting from a full-fledged microscopic description of a given system automatically poses the problem of understanding the emergence of such exponential behavior. This is in general a formidable task, as Lyapunov exponents are emergent, non-perturbative features of the dynamics and there are few physical systems where their existence and dependence on the energy or temperature are known explicitly. For a type of systems satisfying certain homogeneity conditions \cite{seligman_scale-invariant_1985}, the dependence of the classical Lyapunov exponent on the energy can be inferred, and a theory of the quantum Lyapunov exponent, \emph{i.e.,} growth rate, starting from a given dependence on the energy, has been proposed in \cite{hashimoto_bound_2022}. In all these cases, however, the exact form of the results beyond the scaling with the energy is lacking and therefore the closer study of the bound remains out of reach.

So far, no system with a unitary time evolution satisfying these assumptions has been found to violate \cref{eq:mss-bound} to the best of the authors' knowledge, and the cases where the equality is saturated are particularly interesting\footnote{Note however that in non-unitary theories, the bound can be violated \cite{david_chaos_2019}.}. Such systems are usually gravitational, or dual to a gravitational system, and often involve black holes as well\footnote{Beyond the examples already mentioned in the introduction \cite{sachdev_gapless_1993,kitaev_alexei_nodate,noauthor_alexei_nodate,Teitelboim1983,Jackiw1985}, see e.g. \cite{perlmutter_bounding_2016}.}. While it is expected that systems with black holes saturate the bound, there is no consensus on whether a black hole is necessary, or on ``how gravitational'' a system has to be to saturate the bound. An example of such a system is Jackiw-Teitelboim gravity \cite{shenker_multiple_2014} and its dual, the Sachdev-Ye-Kitaev model \cite{Kobrin2020a}. Independent calculations on both sides of the duality show the saturation of the MSS bound in this case. Interestingly, there is a class of gravitational models, the $(2,2p-1)$ minimal string theories, that limit to JT gravity as $p\to\infty$, and for which the question of saturation of \eqref{eq:mss-bound} or not is as yet unanswered \cite{Gregori2021}. We will briefly comment on these models in \cref{sec:leading}.

A more careful analysis of the analytic structure of $F(t)$ that we will touch on in \cref{sec:corrections} reveals that \eqref{eq:corr-leading} are only the first two terms in a short-time expansion of the out-of-time ordered correlator \cite{kundu_subleading_2022},
\begin{equation}
    F_d-F(t)=\epsilon f_1e^{\Lambda t}+\epsilon^2 f_2e^{\Lambda_2t}+\order{\epsilon^3}.
\end{equation}
Here, $\epsilon$ is a small parameter ensuring that the corrections are subleading compared to the leading Lyapunov growth. Note that, consequently, what is referred to in the literature as quantum Lyapunov exponent is \textit{defined} as the rate of growth of the leading contribution to the correlator. In systems such as the one we will introduce in \cref{sec:selberg}, where the scrambling time can be identified with the Ehrenfest time\footnote{When interpreted in a quantum mechanical framework in $f$ dimensions, the dependence of the scrambling time \eqref{eq:scrambling-time} on the entropy $S$ (understood as the number of microstates of volume $\hbar^{f}$) is expected to translate into a logarithmic dependence $t^{*}\sim \beta \log(\hbar)$ in strong analogy with the so-called Ehrenfest time \cite{ehrenfest_bemerkung_1927,berman_condition_1978,rammensee_many-body_2018} $t_{E}\sim\lambda^{-1}\log{(\text{const.}/\hbar)}$, the characteristic time scale that signals the dominance of interference effects in chaotic systems with classical Lyapunov exponent $\lambda$.} $t_E=\lambda^{-1}\log(\text{const.}/\hbar)$ \cite{ehrenfest_bemerkung_1927,berman_condition_1978,rammensee_many-body_2018}, a natural candidate for these subleading terms are quantum corrections, and it is sensible to set $\epsilon=\hbar^2$. 

\section{Wigner-Moyal quantization}\label{sec:wigner-moyal}
\subsection{Basics and definitions}
Given the close, but not yet well-understood connection between classical and quantum Lyapunov exponents, it is helpful to make use of the Wigner-Moyal formalism, which makes the connection between classical (Hamiltonian) and quantum mechanics particularly transparent when post-Ehrenfest time interference effects can be neglected. Being essentially a thermodynamic object, the temperature-dependent quantum Lyapunov exponent is expected to satisfy this condition.

A naive (and for many purposes sufficient) understanding of quantization supposes that one can define a quantum theory by finding operators $X^i,P_i$ for any phase-space degrees of freedom $x^i,p_i$ and identifying commutators with the classical Poisson algebra,
\begin{equation}
    \{x^i,p_j\}=\delta^i_j \xrightarrow{\text{Quantization}}\frac{1}{i\hbar}[X^i,P_j]=\delta^i_j,
\end{equation}
i.e., by taking a ``reverse classical limit''. Trying to generalize this prescription to general phase-space functions, one quickly runs into problems however: Groenewold's theorem states that no quantization map can be found that preserves the classical Poisson structure for all polynomials in $x=(x^1,x^2,\ldots,x^f)$, $p=(p_1,p_2,\ldots,p_f)$ of degree 3 or less \cite{groenewold_principles_1946}.

Moyal subsequently showed that the correct phase-space algebra to represent the quantum operator algebra is not given by Poisson brackets, but by an $\hbar$ deformation of theirs, usually referred to as the Moyal bracket \cite{moyal_quantum_1949}.

To make use of this representation, we need to define the \emph{Weyl symbol} $W_A$ of an operator $A$,
\begin{equation}
    W_A(x,p)=\int dx'\sqrt[4]{g(x+x'/2)g(x-x'/2)}e^{\frac{i}{\hbar}p_ix^i}\mel{x-x'/2}{A}{x+x'/2},\label{eq:weyl-symbol-def}
\end{equation}
with $g(x)$ the determinant of the configuration space metric at the point $x$. Depending on the ordering of the operator $A$, Weyl symbols of different operators with the same classical limit may differ by quantum corrections. 

Some particularly important Weyl symbols are those of phase-space polynomials in the so-called Weyl ordering,
\begin{equation}
    (ax+bp)^n\xrightarrow{\text{Weyl quantization}}(aX+bP)^n.
\end{equation}
In this case, one can just reverse the arrow to find the Weyl symbol,
\begin{equation}
    W_{(aX+bP)^n}=(ax+bp)^n.
\end{equation}
Additionally, we need the Weyl symbol of a density matrix $\rho$, called the Wigner function of the state $\rho$,
\begin{equation}
    W(x,p)\equiv W_\rho(x,p)=\int dx'\sqrt[4]{g(x+x'/2)g(x-x'/2)}e^{\frac{i}{\hbar}p_ix^i}\mel{x-x'/2}{\rho}{x+x'/2}.
\end{equation}
The Wigner function is a quasiprobability distribution on the phase space, and allows for the computation of expectation values in the corresponding state,
\begin{equation}
    \ev{A}_\rho=\int dxdp \,W(x,p)W_A(x,p).
\end{equation}
We further introduce the (Moyal) $\star$ product,
\begin{equation}
    f\star g=\sum_{n=0}^\infty \frac{1}{n!}\left(\frac{i\hbar}{2}\right)^n\Pi^n(f,g),\label{eq:star-product}
\end{equation}
with the Poisson bivector $\Pi=\nabla J\nabla$, where $J=\mqty(0&\mathds{1}\\-\mathds{1}&0)$ is the standard symplectic form\footnote{One can easily see from this definition that $\Pi^{2m+1}(f,g)=-\Pi^{2m+1}(g,f)$, and $\Pi^{2m}(f,g)=\Pi^{2m}(g,f)$.\label{fn:asymmetry}} (on a two-dimensional phase space\footnote{This holds for a phase space with a flat symplectic form $J=\mqty(0&\mathds{1}\\-\mathds{1}&0)$. One can always choose coordinates in which this is locally the case, i.e. for a system with configuration space $\mathcal{X}$, one picks the local trivialization $\mathcal{X}\times\mathds{R}^f$ of the phase space, and then chooses Riemannian normal coordinates on $\mathcal{X}$. If this is not desired, one has to employ the more general Kontsevich quantization formula \cite{kontsevich_deformation_2003}.}):
\begin{equation}
    \begin{aligned}
        \Pi^0(f,g)&=fg,\qquad\Pi^1(f,g)=\{f,g\},\\
        \Pi^n(f,g)=\sum_{k=0}^n&(-1)^k\binom{n}{k}\left(\frac{\partial^k}{\partial p^k}\frac{\partial^{n-k}f}{\partial x^{n-k}}\right)\left(\frac{\partial^{n-k}}{\partial p^{n-k}}\frac{\partial^kg}{\partial x^k}\right),
    \end{aligned}
\end{equation}
and suitably generalized in more dimensions, as well as the Moyal bracket,
\begin{equation}
    \{f,g\}_M=f\star g-g\star f.
\end{equation}
\Cref{eq:star-product} also induces an $\hbar$ expansion in the Moyal bracket,
\begin{equation}
    \{f,g\}_M=\sum_{n=0}^\infty\frac{(-1)^n(\hbar/2)^{2n}}{(2n+1)!}\Pi^{2n+1}(f,g).
\end{equation}
With these definitions, we can use the following properties for Weyl symbols:
\begin{align}
    W_{AB}&=W_A\star W_B,\label{eq:star-product-weyl}\\
    W_{[A,B]}&=i\hbar\{W_A,W_B\}_M.
\end{align}

\subsection{Heisenberg operators}
In order to compute the OTOC, we need a way to find the Weyl symbols of time-evolved (Heisenberg picture) operators. Naively, one could simply cast a Heisenberg operator,
\begin{equation}
    A_t=e^{iHt/\hbar}Ae^{-itH/\hbar},\label{eq:heisenberg-operator}
\end{equation}
in terms of the Weyl symbols of $A$ and the time evolution operator $e^{-iHt/\hbar}$ using \cref{eq:star-product-weyl}. However, the latter exhibits an essential singularity at $\hbar=0$ and therefore does not admit a regular Taylor expansion around that point. Since this is exactly what we want to compute, an alternative way of determining the Weyl symbols of operators like \eqref{eq:heisenberg-operator} is needed.

Osborn and Molzahn \cite{osborn_moyal_1995} provide such a way, which we will closely follow for the remainder of this section. Consider Hamiltonians with a Weyl symbol of the form
\begin{equation}
    W_H(t,\hbar;\zeta)=h_c(t;\zeta)+\sum_{r=1}^\infty\frac{\hbar^r}{r!}h_r(t;\zeta),\label{eq:weyl-hamiltonian}
\end{equation}
where $\zeta=(x,p)$ is an initial-time phase-space point, and the argument $t$ refers to a possible explicit time dependence of the Hamiltonian. We can define the operator $Z$ such that $W_Z=\zeta$, and denote the time evolution from some initial time $s$ to $t$ by 
\begin{equation}
    Z(t,s)=\Gamma(s,t)Z,\quad H(t,s)=\Gamma(s,t)H(t).
\end{equation}
The time evolution is then determined by the Heisenberg equation,
\begin{equation}
    i\hbar\frac{d}{dt}Z(t,s)=[Z(t,s),H(t,s)]=\Gamma(s,t)[Z,H(t)]. \label{eq:heisenberg-equation}
\end{equation}
The second equation usefully reveals the dependence of the so-called quantum trajectory $Z(t,s)$ on the commutator between the initial condition $Z$ and the Hamiltonian. \Cref{eq:heisenberg-equation} is solved by the ansatz
\begin{equation}
    W_Z(t,s,\hbar;\zeta)=\sum_{r=0}^\infty\frac{\hbar^r}{r!}z_r(t,s;\zeta),\label{eq:ansatz}
\end{equation}
where the expansion coefficients are determined as follows:

The classical limit of \cref{eq:heisenberg-equation} is 
\begin{equation}
    \frac{d}{dt}z_0(t,s;\zeta)=J\nabla h_c(t,z_0(t,s;\zeta)),
\end{equation}
which is precisely Hamilton's equation and, therefore, is solved by the classical flow generated by the Hamiltonian $h_c$,
\begin{equation}
    z_0(t,s;\cdot)=\gamma(t,s\vert\cdot),
\end{equation}
i.e. $\gamma(t,s\vert(x_s,p_s))=(x_t,p_t)$, or simply, the classical solution of the equations of motion. For higher order coefficients, we define the Jacobi operator
\begin{equation}
    \mathcal{J}(t;s,\zeta)=\frac{d}{dt}-J\nabla\nabla h_c(t,\gamma(t,s\vert\zeta)),
\end{equation}
which gives equations of the form
\begin{equation}
    \mathcal{J}(t;s,\zeta)z_r=f_r(t,s;\zeta).\label{eq:zr-deq}
\end{equation}
In the following, we will only consider the first
\begin{equation}
    \mathcal{J}(t;s,\zeta)z_1(t,s;\zeta)=J\nabla h_1(t,\gamma(t,s\vert\zeta)),\label{eq:z1-deq}
\end{equation}
and second equation\footnote{Expressions such as $\Pi_{ij}f_1\ldots f_n$ are to be understood as taking derivatives w.r.t. the arguments $i$ and $j$, then evaluating at $\zeta_k=\zeta$ for all $k=1,\ldots,n$.},
\begin{equation}
    \begin{aligned}
        \mathcal{J}(t;s,\zeta)z_2(t,s;\zeta)=&\left[(z_1\cdot\nabla)^2-\frac{1}{8}\Pi^2(\gamma\cdot\nabla)^2+\frac{1}{12}\Pi_{12}\Pi_{23}(\gamma\cdot\nabla)^3\right]J\nabla h_c(t,\gamma(t,s\vert\zeta))\\[1em]
        &+2(z_1\cdot\nabla)J\nabla h_1(t,\gamma(t,s\vert\zeta))+J\nabla h_2(t,\gamma(t,s\vert\zeta)).\label{eq:z2-deq}
    \end{aligned}
\end{equation}
These equations can then be integrated using a Green's function of the Jacobi operator $\mathcal{J}$, which can be chosen as a function of derivatives of the classical flow $\gamma$ \cite{osborn_moyal_1995},
\begin{equation}
    z_r(t;\zeta)=\int_0^t ds\nabla\gamma(t,0\vert\zeta)J\nabla\gamma(s,0\vert\zeta)^TJ^{-1}f_r(s,0;\zeta).\label{eq:zr-integrated}
\end{equation} 
A similar treatment allows for finding the Weyl symbols of more complicated operators, but in this work, we shall only be concerned with quantum trajectories $Z(t,s)$. A nice side effect of this method is that it does not make explicit reference to the potentially curved configuration space: any difficulties arising in that context are encoded in the classical flow $\gamma$, and in the Weyl symbol of the Hamiltonian \eqref{eq:weyl-hamiltonian}.

\section{Particle on a high-dimensional hyperbolic manifold}
\label{sec:selberg}

In this section, we will introduce a model in which to compute the OTOC using the formalism developed in \cref{sec:wigner-moyal}. Since we want to examine Lyapunov growth, a sensible requirement is that the dynamical system be chaotic, and the ``more chaotic'', the better, if we want to get close to the MSS bound. An interesting possibility is to use a Hadamard-Gutzwiller-like model (e.g. \cite{aurich_periodic_1988,Aurich1989}), that is, a particle of mass $m$ moving freely on a surface $\mathcal{M}$ of constant curvature $R=-2/L^2$.
\begin{figure}
    \centering
    \includegraphics[width=0.6\linewidth]{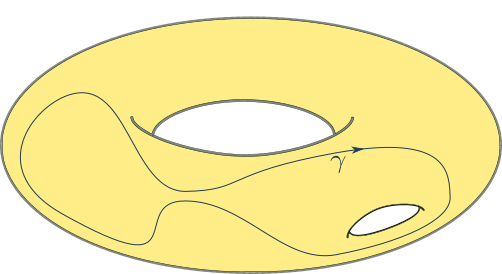}
    \caption{Example surface for a Hadamard-Gutzwiller-like model. The blue curve is a possible periodic orbit of the classical dynamics. In order for the motion to be chaotic, the surface has to have genus at least 2. Adapted from \cite{haneder_beyond_2025}.}
    \label{fig:system}
\end{figure}
An example of such a system is sketched in \cref{fig:system}, together with a periodic orbit of the classical dynamics, generated by the Hamilton function
\begin{equation}
    h=\frac{1}{2m}p_ig^{ij}p_j,
\end{equation}
where $p_{i=1,2}$ are the momenta canonically conjugate to coordinates on the Poincar\'{e} disk $\mathds{H}_2$, $x^{i=1,2}$, and $g^{ij}$ is the inverse of the metric
\begin{equation}
    g_{ij}=\frac{4L^2}{(1-(x^1)^2-(x^2)^2)^2}\delta_{ij}.
\end{equation}
Surfaces like the one depicted in \cref{fig:system} can be realized as a quotient of $\mathds{H}_2$ with some discrete group of isometries $\Gamma$. After canonical quantization, the Hamiltonian of such a system reads
\begin{equation}
    H=-\frac{\hbar^2}{2mL^2}\Delta,\label{eq:hadamard-gutzwiller-hamiltonian}
\end{equation}
where $\Delta$ is the (dimensionless\footnote{In two dimensions, any compact surface that is neither isomorphic to a sphere, nor a torus, can be endowed with a metric of constant curvature $R=-2$ by the uniformization theorem \cite{poincare_sur_1908}. $\Delta$ is the Laplace-Beltrami operator computed from this metric.}) Laplace-Beltrami operator on $\mathcal{M}$, and $\frac{\hbar^2}{2mL^2}$ sets the energy scale of the system.

A very special feature of this type of system is that the full quantum system is completely equivalent to its semiclassical description, due to the presence of the Selberg trace formula (STF) \cite{selberg_harmonic_1956},
\begin{equation}
    \sum_nu\left(\sqrt{\frac{2mL^2}{\hbar^2}E_n}\right)=\frac{L^2V}{4\pi}\int_0^\infty dr\, u(r)\Phi_2(r)+\sum_{\text{PO}}\sum_{k=1}^\infty A_{\text{PO}}\tilde{u}\left(\frac{kl_{\text{PO}}}{L}\right).\label{eq:hadamard-gutzwiller-stf}
\end{equation}
Here, $u(r)$ is a function of the spectrum of the Laplacian, $E_n$ are the eigenenergies of \eqref{eq:hadamard-gutzwiller-hamiltonian}, $L^2V$ is the area of $\mathcal{M}$, $\Phi_2(r)$ is the so-called \emph{Plancherel measure} of $PSL(2,\mathds{R})$ (the isometry group of $\mathds{H}_2$), the sum $\sum_{\text{PO}}$ ranges over primitive classical periodic orbits of the system, while the sum $\sum_k$ counts their repetitions, $A_{\text{PO}}$ is the stability amplitude of a given periodic orbit, $l_{\text{PO}}$ its length, and finally $\tilde{u}(l/L)$ is the Fourier transform of the spectral function $u(r)$.

\Cref{eq:hadamard-gutzwiller-stf} is, for the system at hand, equivalent to the Gutzwiller trace formula \cite{gutzwiller_periodic_1971} widely used in semiclassical physics and periodic-orbit theory. Usually, Gutzwiller's trace formula is a small $\hbar$ approximation to the full quantum path integral, but \cref{eq:hadamard-gutzwiller-stf} is an exact mathematical identity.

For the STF to apply, the spectral function $u(r)$ has to satisfy certain conditions, one of which, interestingly, is an analyticity condition: it must be analytic in a strip of width greater than 1 around the real axis. This condition is reminiscent of (but not equivalent to) the condition on the analyticity of the OTOC in \cite{Maldacena2016} that is required to derive the MSS bound on the Lyapunov exponent \cite{minar_periodic_2024}.

We can also generalize this system to higher (indeed, arbitrarily high) dimensions; a single particle in such a high-dimensional configuration space can represent a many-body system with many degrees of freedom. In this setting, the classical dynamics (and correspondingly, any canonical quantum description) may become more complicated, but the group theoretical construction underpinning \cref{eq:hadamard-gutzwiller-stf} generalizes straightforwardly, and one can obtain a Selberg trace formula in arbitrary dimension $f$ \cite{randol_selberg_1984,bytsenko_quantum_1996},
\begin{equation}
    \sum_nu\left(\sqrt{\frac{2mL^2}{\hbar^2}E_n}\right)=\frac{L^fV}{(4\pi)^{f/2}}\int_0^\infty dr\, u(r)\Phi_f(r)+\sum_{\text{PO}}\sum_{k=1}^\infty A_{\text{PO}}\,\tilde{u}\left(\frac{kl_{\text{PO}}}{L}\right).\label{eq:stf}
\end{equation}
The analyticity condition mentioned above has to be modified slightly; the strip where $u$ is analytic now has to have a width of at least $f-1$, and the manifold on which the system lives is now $\mathcal{M}=\mathds{H}_d/\Gamma$, with $\Gamma$ again a discrete group of isometries. \Cref{eq:stf} has recently been used to reproduce correlation functions of JT gravity \cite{haneder_beyond_2025} in the formal limit $f\to\infty$, and as such, this kind of system makes for an interesting candidate for the study of OTOCs, particularly as relates to the MSS bound. 

The STF is not particularly useful for the computation of the OTOC we intend to perform (although there are examples of OTOC calculations using semiclassical theory, e.g. \cite{rammensee_many-body_2018,jalabert_semiclassical_2018}), but it reveals one of the ingredients needed for our computation: the Plancherel measure $\Phi_f(r)$. After substituting $r=\sqrt{\frac{2mL^2E}{\hbar^2}}$,
\begin{equation}
    \Phi_f(r)dr\equiv\varrho_f\left(E\right)\sqrt{\frac{2mL^2}{\hbar^2 E}}dE,
\end{equation}
where $\varrho_f(E)$ denotes the density of states, simply counts the number of states in an energy interval. It therefore doubles as the microcanonical partition function of the theory, which will be important in \cref{sec:leading}. The Plancherel measure is known in the literature (after absorbing the Jacobi determinant factor),
\begin{equation}
    \begin{aligned}
    \varrho_f(E)&=\sqrt{\frac{2mL^2}{\hbar^2 E}}\frac{f}{(4\pi)^{f/2}\Gamma\left(\frac{f+2}{2}\right)}\frac{\abs{\Gamma\left(i\sqrt{\frac{2mL^2}{\hbar^2}E}+(f-1)/2\right)}^2}{\abs{\Gamma\left(i\sqrt{\frac{2mL^2}{\hbar^2}E}\right)}^2}\\
    &=
    \begin{dcases}
        \frac{2mL^2}{\hbar^2}\frac{\tanh(\pi \sqrt{\frac{2mL^2}{\hbar^2}E})}{(2\pi)^{f/2}(f-2)!!}\prod_{k=0}^{\frac{f-4}{2}}\left(\frac{2mL^2}{\hbar}E+\left(k+\frac{1}{2}\right)^2\right) & f\text{~even}\\
        \frac{1}{2^{(f-1)/2}\pi^{(f+1)/2}(f-2)!!}\sqrt{\frac{2mL^2}{\hbar^2 E}}\prod_{k=0}^{\frac{f-3}{2}}\left(\frac{2mL^2}{\hbar^2}E+k^2\right) & f\text{~odd}.
    \end{dcases}
    \end{aligned}
    \label{eq:rho}
\end{equation}
Most notably, in the limit $f\to\infty$, the density of states reads \cite{haneder_beyond_2025}
\begin{equation}
    \varrho_\infty(E)=\frac{2mL^2}{\hbar^2}\sinh\left(\pi\sqrt{\frac{2mL^2}{\hbar^2}E}\right),\label{eq:rho-infinity}
\end{equation}
{\textit i.e.}, it is (up to a rescaling) equal to the $\sinh$ or Schwarzian density of states characteristic of $2d$ dilaton gravity \cite{mertens_solvable_2023}.

As we will show in the next section, the semiclassical calculation of the quantum Lyapunov exponent will critically depend not only on the precise knowledge of the level density, \cref{eq:rho}, but also on the specific form of the classical, microcanonical Lyapunov exponent. Remarkably, in our case, the latter is not only fully independent of the dimension\footnote{In generic chaotic systems with $f$ degrees of freedom, one would expect $f$ possibly different positive classical Lyapunov exponents. In the system at hand however, all positive Lyapunov exponents are equal \cite{butler_rigidity_2015}.}, but its rigorously exact dependence on the microscopic parameters of the theory and the energy is known and given by \cite{aurich_periodic_1988}

\begin{equation}
    \lambda(E)=\sqrt{\frac{2E}{mL^2}},\label{eq:classical-lyapunov}
\end{equation}
where $E=p^2/2m$. \Cref{eq:classical-lyapunov} can be understood heuristically by noticing that it is essentially a measure of the curvature pushing geodesics away from each other. This heuristic picture can be made rigorous by means of the solutions of the corresponding Jacobi fields \cite{gutzwiller_chaos_1990,KobayashiNomizu1969}.

Equipped with these two ingredients, the exact mean level density and the classical microcanonical Lyapunov exponent, we can proceed to the leading order computation of the quantum Lyapunov exponent.

\section{Leading-order quantum Lyapunov exponent}\label{sec:leading}
\subsection{General strategy}

We start with the computation of the OTOC for systems with classical chaotic limit, up to leading order in $\hbar$ with the operator choice $V=X, W=P$:

\begin{equation}
    \mathcal{C}(t)=-\tr\left(\rho(\beta)[P_t,X]^2\right),
\end{equation}
where $\rho(\beta)=\frac{e^{-\beta H}}{Z(\beta)}$ is the thermal state, as specified in \cref{sec:otoc}. In exchange for introducing a factor $\delta(E-H)$, we can rewrite this expression in terms of an energy integral. The result can then be related to the microcanonical average of the commutator under consideration:

\begin{align}
    \mathcal{C}(t)&=-\int_0^\infty dE\tr\left(\frac{e^{-\beta H}}{Z(\beta)}[P_t,X]^2\delta(E-H)\right)\nonumber\\
    &=-\int_0^\infty dE \frac{e^{-\beta E}}{Z(\beta)}Z(E)\frac{1}{Z(E)}\tr\left(\delta(E-H)[P_t,X]^2\right)\nonumber\\
    &=-\int_0^\infty dE\frac{Z(E)}{Z(\beta)}e^{-\beta E}\ev{[P_t,X]^2}_{\text{mc}},
\end{align}
where we used the fact that $\delta(E-H)$ is precisely the microcanonical density matrix, and $Z(E)=\tr\delta(E-H)$. We now express the microcanonical average through its Wigner-Moyal quantization,
\begin{equation}
    \mathcal{C}(t)=-\int_0^\infty dE\frac{Z(E)}{Z(\beta)}e^{-\beta E}\int dxdp \,W_{[P_t,X]^2}W(x,p),\label{eq:otoc_weyl_symbs}
\end{equation}
where $(x,p)\in\mathds{R}^{2f}$ are coordinates parametrizing the phase space of the system, $W_{A}$ is the Weyl symbol of the operator $A$ and $W(x,p)=W_{\delta(E-H(X,P))}$ is the Wigner function. For simplicity of notation, we will assume that the phase space is covered by a single coordinate patch, so that there is only one region that contributes to the phase-space integral in \cref{eq:otoc_weyl_symbs}\footnote{A further subtlety arises from the fact that on compact manifolds, momenta conjugate to coordinates with finite range (e.g. angles) are quantized, leading to a discrete phase space. However, after replacing the corresponding momentum integrals by sums, our arguments are still applicable. In the semiclassical limit, the continuous phase space can be recovered.}. 

Directly computing the Weyl symbol of such a complicated operator is quite hard, but we can simplify the task somewhat:
\begin{align}
    W_{[P_t,X]^2}&=W_{[P_t,X]}\star W_{[P_t,X]}\nonumber\\
    &=-\hbar^2\{W_{P_t},W_{X}\}_M\star\{W_{P_t},W_{X}\}_M\label{eq:otoc-weyl}
\end{align}
The Weyl symbol of Weyl ordered initial time operators is simply given by the direct replacement of all factors $X,P\to x,p$, i.e. by their corresponding phase-space functions. For time-evolved or not Weyl-ordered operators, finding the Weyl symbol is much more involved (as we will see in \cref{sec:corrections}), but the beauty of the Wigner-Moyal approach lies in the fact that at leading order in $\hbar$, we can simply replace
\begin{enumerate}
    \item the Weyl symbol $W_{P_t}$ by the classical solution $p_t$,
    \item the Moyal bracket by the Poisson bracket and
    \item the $\star$ product by the usual phase-space product.
\end{enumerate}
Likewise, at leading order, the Wigner function is given by
\begin{equation}
    W_E(x,p)=\frac{\delta(E-h(x,p))}{Z(h(x,p))}+\order{\hbar}.\label{eq:wigner-function-leading}
\end{equation}
A crucial observation is that, for the OTOC understood as an expansion in a small parameter \cite{Maldacena2016} with corrections becoming important at later times \cite{kundu_subleading_2022,kundu_extremal_2022}, our calculation up to this point shows that the leading order Lyapunov exponent is \emph{independent} of the regularization (i.e. choice of distribution of the density matrix factors) -- a property so far only assumed and confirmed numerically, but never rigorously shown \cite{tsuji_bound_2018}.

After the above replacements, the OTOC reads, to leading order,
\begin{equation}
\label{eq:otocgen}
    \mathcal{C}(t)\approx\hbar^2\int_0^\infty dE\, \frac{Z(E)}{Z(\beta)}e^{-\beta E}\int dxdp\, W_E(x,p)\abs{\pdv{p^i_t(x,p)}{p^j}}^2,
\end{equation}
where we made the dependence of $p_t$ on the initial conditions explicit, and we use the Einstein convention to indicate summation over the indices $i,j=1,\ldots,f$. Since we are only interested in the growth rate of this integral, evaluating it exactly is not necessary. Now the classical microcanonical Lyapunov exponent $\lambda$ enters, capturing the exponential growth of the off-diagonal blocks $ \{p_t^i,q_j\}$,

\begin{equation}
\label{eq:poisson}
    \{p_t^i,q_j\}=-\pdv{p^i_t(x,p)}{p^j}=F^i_j(x,p) e^{\lambda(x,p)t},
\end{equation}
of the stability matrix \cite{gaspard_chaos_1998}.

Up to now, our only assumption is that the system is chaotic and admits a well-defined classical limit, with a leading-order approximation in $\hbar$ to the OTOC given by inserting \cref{eq:poisson} into \cref{eq:otocgen}. The canonical quantum Lyapunov exponent is then obtained by identifying the corresponding leading order-contribution in $\hbar$ to the exponential growth (if any) of the OTOC. Progress in this direction is only possible if the specific dependences of both the thermodynamic $Z(E),Z(\beta)$ and dynamical $\lambda(x,p)$ functions entering \cref{eq:otocgen} are known.

\subsection{The canonical quantum Lyapunov exponent for high-dimensional hyperbolic motion}

To apply the method above to high-dimensional hyperbolic motion, we focus on systems where the Lyapunov exponent is constant on the classical energy shell, namely where $\lambda(x,p)$ depends on the initial conditions $(x,p)$ only through the energy $E=h(x,p)$, such as the ones considered in \cite{hashimoto_bound_2022} and \cite{seligman_scale-invariant_1985,gaspard_chaos_1998}. This is a special and convenient feature, as the dependence of $\lambda(x,p)$ on the initial phase space region in general systems is expected to be complicated and, in most cases, simply not known. $F^i_j(x,p)$ is a slowly varying phase-space function that will not be important for our purposes\footnote{Roughly, $F^i_j(x,p)$ is a function of the basis vectors of the tangent space at the phase-space point $(x,p)$, while the exponential behavior is captured by the stretching factor $e^{\lambda t}$ \cite{gaspard_chaos_1998}.}. Using \cref{eq:wigner-function-leading}, we have
\begin{equation}
    W_E(x,p)e^{2\lambda(h(x,p))t}=W_E(x,p)e^{2\lambda(E)t}+\order{\hbar},
\end{equation}
so we can pull the exponential term out of the phase-space integral and are left with
\begin{align}
    \mathcal{C}(t)&=\hbar^2\int_0^\infty dE\frac{Z(E)}{Z(\beta)}e^{2\lambda(E)t-\beta E}\,\overline{\abs{F^i_j}^2}(E)\nonumber\\
    &=\frac{\hbar^2}{Z(\beta)}\int_0^\infty dE\,e^{2\lambda(E)t-\beta E+\log Z(E)}\,\overline{\abs{F^i_j}^2}(E)\label{eq:preSPA}
\end{align}
where $\overline{\abs{F^i_j}^2}(E)$ is the phase-space average of the slowly varying part of the stability matrix, and hence is expected to only weakly depend on the energy as well. 

When $f\!=\!2$, this result for the leading $\order{\hbar^2}$ contribution to the OTOC coincides with the result of \cite{jalabert_semiclassical_2018} for the appropriate choice of operators. However, it does not allow for an unambiguous identification of a quantum Lyapunov exponent as there is no clear region of exponential growth. As we show below, this is because the key ingredient for such an identification is a saddle point analysis only justified in the regime of ensemble equivalence, $f\! \to \!\infty$, where the bound on chaos was originally derived \cite{Maldacena2016}.

We will invoke the standard tools of ensemble equivalence well known in statistical physics \cite{martin_theory_1959}, that are asymptotically exact in the limit $f\to \infty$. From \cref{eq:preSPA}, the growth rate of the integral, and thereby the Lyapunov exponent, can be estimated (up to loop corrections) by evaluation at the stationary point $E_\beta^*$ of the integrand,
\begin{equation}
    \mathcal{C}(t)\approx\hbar^2\overline{\abs{F^i_j}^2}(E_\beta^*)e^{2\lambda(E_\beta^*)t},\label{eq:saddle-point-otoc}
\end{equation}
where we used that $Z(\beta)=Z(E_\beta^*)e^{-\beta E_\beta^*}$ \cite{martin_theory_1959,horing_quantum_2017} as dictated by the standard thermodynamic relation between the entropy and the free energy. Plugging in the microcanonical Lyapunov exponent \eqref{eq:classical-lyapunov} of the system, the stationarity condition reads
\begin{equation}
    0=2\lambda'(E_\beta^*)t-\beta+\frac{Z'(E_\beta^*)}{Z(E_\beta^*)}\equiv\frac{\sqrt{2}t}{\sqrt{mL^2E_\beta^*}}-\beta+G_f(\beta),\label{eq:stationarity}
\end{equation}
where $G_f(\beta)=\varrho'_f(E)/\varrho_f(E)$ is determined by approximating the microcanonical partition function $Z(E)$ by its smooth part, $Z(E)\approx\varrho_f(E)$.

We can analytically evaluate \cref{eq:stationarity} in two interesting regimes. The density of states \eqref{eq:rho} of our model (we may restrict to odd $f$ for simplicity) takes the form of a curvature expansion \cite{Balian1970a,Balian1971,Balian1972}. This expansion comprises the highest-degree monomial given by the Weyl volume law \cite{Haake2010a}, as well as all quantum corrections in the form of a lower-degree polynomial \cite{Balazs1986}. Therefore, if we want to access the classical regime of the system, the highest-degree term in \cref{eq:rho} should dominate the others \cite{sethna_statistical_2021}. We can translate this to the simple condition (see appendix \ref{app:regime} for a short derivation),
\begin{equation}
    \frac{f^2}{24}\ll\frac{2mL^2}{\hbar^2\beta}\equiv\frac{4\pi L^2}{\lambda_{\mathrm{th}}^2},\label{eq:classical-condition}
\end{equation}
where we introduced the particle's thermal de Broglie wavelength,
\begin{equation}
    \lambda_{\text{th}}=\sqrt{\frac{2\pi\hbar^2\beta}{m}}.
\end{equation}
If the number of degrees of freedom $f$ is large, which is necessary for the applicability of the saddle point approximation \eqref{eq:saddle-point-otoc}, the condition \eqref{eq:classical-condition} simply means that their thermal wavelength has to be small enough not to experience curvature effects. Since curvature corrections to the density of states play the role of quantum corrections in our system, this is consistent with the aim of studying the system in the classical regime. Neglecting the quantum corrections then, we can solve the stationarity condition \eqref{eq:stationarity} with

\begin{equation}
    G_f(\beta)=\frac{f}{2E_\beta^*}
\end{equation}
and find the growth rate of the OTOC to be
\begin{equation}
    2\lambda(E_\beta^*)=\frac{\sqrt{\frac{4t^2}{mL^2}}\pm\sqrt{\frac{4t^2}{mL^2}+4\beta f}}{\sqrt{mL^2\beta^2}},
    \label{eq:small-dim-lyapunov-full}
\end{equation}
which is independent of $\hbar$, as expected. If we assume that the thermal energy per degree of freedom is much smaller than the kinetic energy of a particle that is sensitive to curvature effects on a timescale $t$,
\begin{equation}
    \frac{1}{\beta}\ll \frac{mL^2}{t^2},
\end{equation}
we can neglect the $t$-dependent terms in \cref{eq:small-dim-lyapunov-full} and find

\begin{equation}
    2\lambda(E_\beta^*)=\sqrt{\frac{4f}{mL^2\beta}}.\label{eq:small-dim-lyapunov}
\end{equation}
Recalling \cref{eq:classical-lyapunov}, this is simply the statement that the energy in the system is given by the classical equipartition theorem. Usually, one would associate larger, rather than smaller temperatures to more classical behavior, but in the system at hand, it is known that quantum corrections to the density of states take the form of a curvature expansion \cite{Balian1970a,Balian1971,Balian1972}, and hence, it is sensible to stay far away from the regime in which such corrections start mattering to observe the system at its ``most classical''. 

As the thermal wavelength of the particle grows to comparable size to the curvature radius, i.e. as the temperature decreases, quantum corrections become very important, and eventually, the entire polynomial \eqref{eq:rho} will contribute, approximating the infinite-dimensional $\sinh$-behavior. It is instructive to exploit the product structure of \cref{eq:rho} and write the density of states (again in the $f$ odd case for simplicity) as the $f\to\infty$ result up to multiplicative corrections,
\begin{equation}
    \rho_{f}(E)\propto\frac{\sinh(\pi\sqrt{\frac{2mL^2}{\hbar^2}E})}{\prod_{k=\frac{f-1}{2}}^\infty k^2+\frac{2mL^2}{\hbar^2}E},\label{eq:high-dim-rho}
\end{equation}
which yields
\begin{equation}
    G_f(\beta)=\sqrt{\frac{mL^2}{2\hbar^2}}\left(\frac{\pi}{\sqrt{E_\beta^*}}\coth(\pi\sqrt{\frac{2mL^2}{\hbar^2}E_\beta^*})-\sum_{k=1}^{\infty}\frac{1}{(k+f)^2/4+2mL^2E_\beta^*/\hbar^2}\right).
\end{equation}
Using the fact that
\begin{equation}
    \coth{x}=\frac{1+e^{-2x}}{1-e^{-2x}}\approx1\quad\mathrm{for~}x\gtrsim1,
\end{equation}
we can neglect the $\coth$ factor whenever
\begin{equation}
    E_\beta^*\gtrsim\frac{\hbar^2}{2mL^2}\frac{1}{\pi^2}.\label{eq:coth-condition}
\end{equation}
With this simplification, it remains to solve the stationarity condition \eqref{eq:stationarity} for
\begin{equation}
    G_f(\beta)\approx\sqrt{\frac{mL^2}{2\hbar^2}}\left(\frac{\pi}{\sqrt{E_\beta^*}}-\sum_{k=1}^{\infty}\frac{1}{(k+f)^2/4+2mL^2E_\beta^*/\hbar^2}\right).\label{eq:saddle-G-large}
\end{equation}

\begin{figure}
    \centering
    \includegraphics[width=0.9\linewidth]{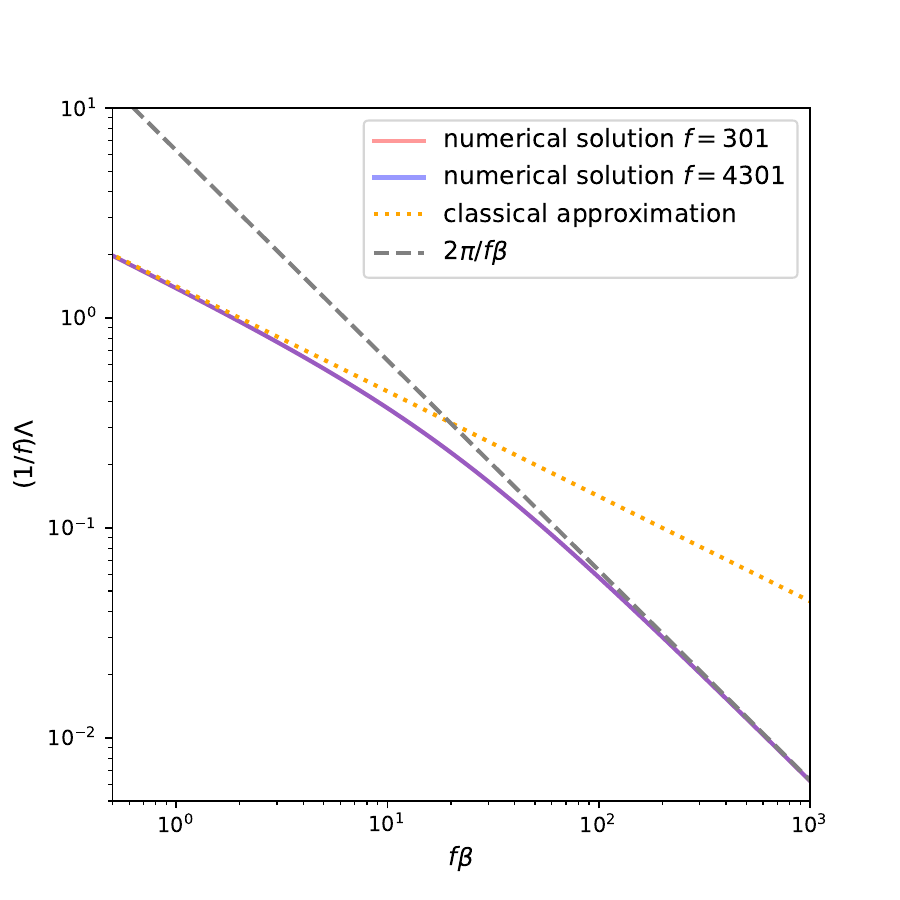}
    \caption{Double-$ \log$ plot of the system's rescaled leading-order quantum Lyapunov exponent $(1/f)\Lambda$ (where $\Lambda=2\lambda(E_\beta^*)$) as a function of the rescaled inverse temperature $f\beta$. The slightly opaque solid lines show $(1/f)\Lambda(f\beta)$, obtained from the full numerical solution of the stationarity condition \eqref{eq:stationarity} for two representative dimensions, $f\!=\!301$ and $f\!=\!4301$, in units where $\hbar=1,m=2,L=1$. 
   The two numerical curves are so close to each other that their difference cannot be resolved in the plot.
  This indicates $(f\beta)$-scale invariance in the large-$f$ limit. 
   Dotted orange line: classical approximation to the Lyapunov exponent \eqref{eq:small-dim-lyapunov}, given by the equipartition theorem. Dashed grey line: $2\pi/\beta$, i.e. the MSS bound eq.~(\ref{eq:infinite-lyapunov}). The system saturates the MSS bound at low (scaled) temperatures, while it is more appropriately described by classical equipartition  (eq.~\eqref{eq:small-dim-lyapunov}) at high (scaled) temperatures.}
    \label{fig:mss-convergence-temperature}
\end{figure}

If the dimension is sufficiently large, the sum in \cref{eq:saddle-G-large} should only result in a small correction to the stationarity condition \eqref{eq:stationarity},
\begin{equation}
    \frac{f^2}{4}\gg\frac{2mL^2E_\beta^*}{\hbar^2}\approx\frac{4L^4}{\lambda_\text{th}^4},
\end{equation}
where the ``$\approx$'' comes from plugging in the solution 
\begin{equation}
E_\beta^*=\frac{mL^2\pi^2}{\hbar^2\beta^2} \, , 
 \label{eq:E_ast}
\end{equation}
obtained from solving the saddle point equation neglecting the sum in \cref{eq:saddle-G-large}\footnote{This energy has to be in particular consistent with \cref{eq:coth-condition}.}. Note that this describes (roughly) the opposite extreme to the classicality condition \eqref{eq:classical-condition}, i.e. at fixed dimension, the thermal wavelength needs to increase until curvature (and thus quantum) effects become relevant to access it. This justifies solving \cref{eq:stationarity} iteratively. Neglecting the last term, solving for $E_\beta^*$, see eq.~(\ref{eq:E_ast}), 
inserting this again into the full equation, and solving a second time, results in
\begin{equation}
\begin{aligned}
    2\lambda(E_\beta^*)=&\frac{2(\hbar t+\pi mL^2)}{mL^2\hbar \beta}\frac{1}{1+\frac{2}{\pi}\Im(\Psi (1+f-4i\frac{\hbar t+\pi mL^2}{\hbar^2\beta}))}\\&\xrightarrow{\hbar\ll \frac{mL^2}{t}}\frac{2\pi}{\hbar\beta}\left(1-\frac{16\pi L^2}{\lambda_{\text{th}}^2}\frac{\log(f+1)}{f+1}+\dots\right),\label{eq:large-dim-lyapunov}
    \end{aligned}
\end{equation}
with the digamma function $\Psi(z)\!=\!\frac{\Gamma'(z)}{\Gamma(z)}$. The regime indicated by the arrow is accessed if the typical action of a trajectory of the particle $\frac{mL^2}{t}$ is large against $\hbar$, i.e. when our initially very quantum system becomes ``more classical'' again\footnote{It should be noted at this point that particularly in many-body semiclassics, there are typically two complementary notions of the classical limit \cite{richter_semiclassical_2022}: the usual classical limit $\hbar\to0$, as well as the limit of a vanishing ``effective'' $\hbar$, i.e. the number of degrees of freedom of the system $f\to\infty$. It has been argued for bosonic systems in \cite{rammensee_many-body_2018} that the limit $f\to\infty$ produces an expansion of the OTOC akin to the one of \cite{Maldacena2016}, in apparent tension with our results. A key difference between the case of \cite{rammensee_many-body_2018} and ours, however, is that the degrees of freedom described by \cref{eq:hadamard-gutzwiller-hamiltonian} are distinguishable, calling into question the simple applicability of results for bosonic systems. Indeed, the limit $f\to\infty$ is not a natural classical limit in our case; it is rather the limit that emphasizes the quantum regime \emph{the most}, in the sense of making the system maximally chaotic at every temperature, see \cref{fig:mss-convergence-temperature}.}. 
From here, we can see the correction coming from the sum in \cref{eq:saddle-G-large} dying away if the thermal wavelength increases, or alternatively, if the dimension becomes very large. Indeed, in the infinite-dimensional limit, the sum in \cref{eq:saddle-G-large} vanishes entirely, and we are only left with
\begin{equation}
    G_f(\beta)\approx
        \sqrt{\frac{2mL^2}{\hbar^2}}\frac{\pi}{\sqrt{E_\beta^*}},\quad f=\infty,
\end{equation}
which yields for the Lyapunov exponent the central result
\begin{equation}
    2\lambda(E_\beta^*)=\frac{2}{\hbar}\left(\frac{\pi}{\beta}+\frac{\hbar}{mL^2}\frac{t}{\beta}\right) \xrightarrow{\hbar\ll\frac{mL^2}{t}}\frac{2\pi}{\hbar\beta}.\label{eq:infinite-lyapunov}
\end{equation}

\begin{figure}
    \centering
    \includegraphics[width=0.9\linewidth]{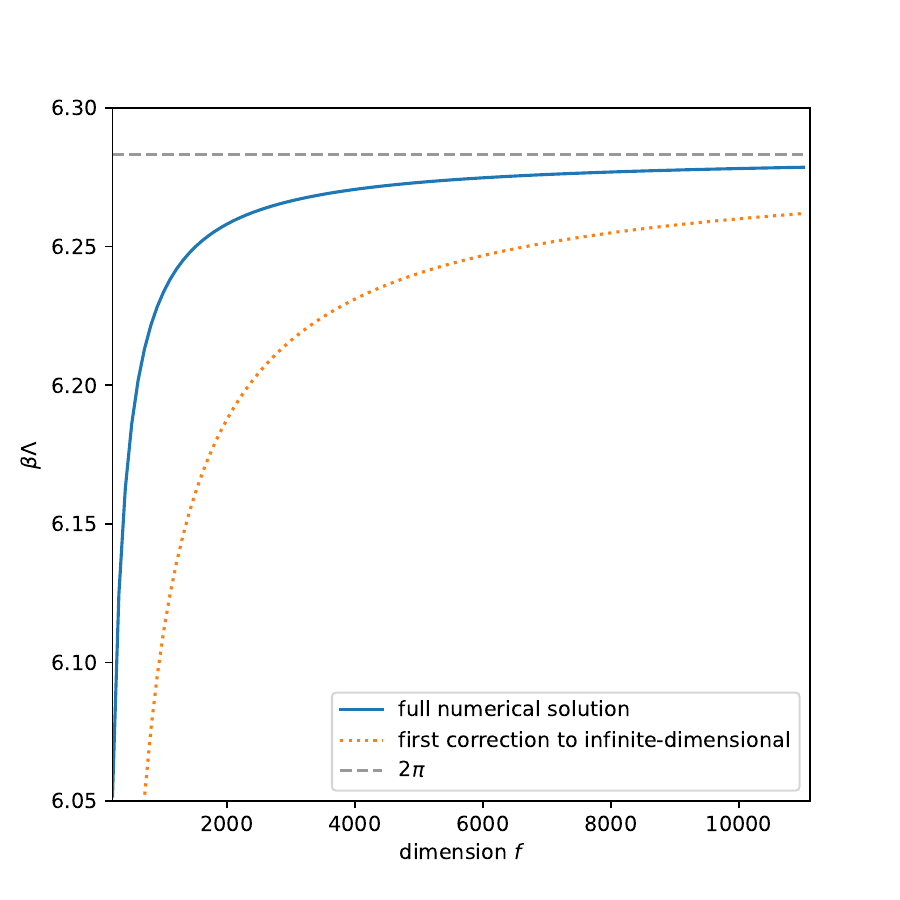}
    \caption{Quantum Lyapunov exponent $\Lambda=2\lambda(E_\beta^*)$ for inverse temperature $\beta=1$, evaluated for different dimensions $f$, plotted in units where $\hbar=1,m=2,L=1,\lambda_{\mathrm{th}}=1/2$. Solid blue line: full numerical solution of the stationarity condition \eqref{eq:stationarity}. Dotted orange line: first correction to the infinite-dimensional solution for large, finite dimension \eqref{eq:large-dim-lyapunov}. Grey dashed line: $2\pi$, i.e. the MSS bound, eq.~(\ref{eq:mss-bound}). One can see that both finite-dimensional results approach the MSS bound for (very) large dimensionality.}
    \label{fig:mss-convergence-dimension}
\end{figure}

Remarkably, our chaotic quantum system saturates the MSS bound, corresponding to maximally fast scrambling, in the limit of infinite configuration space dimension\footnote{In this context, the conjecture proposed in \cite{hashimoto_bound_2022} can be interpreted as a \textit{classical} bound, that is corrected by increasingly strong quantum corrections, eventually producing the \textit{quantum} bound, with a transition that happens around the $f$-dependent crossing point seen in \cref{fig:mss-convergence-temperature}.}.
Since this is precisely the limit in which it starts to exhibit correlation functions akin to the ones found in JT gravity \cite{haneder_beyond_2025}, this result serves to further support the status of this model as dual to JT gravity, where saturation of the MSS bound has been confirmed independently \cite{shenker_multiple_2014}. The approach of the system's quantum Lyapunov exponent to the bound by increasing the dimension can be seen in \cref{fig:mss-convergence-dimension}, where we plot the full solution of the stationarity condition \eqref{eq:stationarity}, as well as the first correction to the infinite-dimensional result \eqref{eq:large-dim-lyapunov}. In both cases, we can see that the MSS bound is approached\ when increasing the dimension $f$.

\Cref{fig:mss-convergence-temperature} shows the quantum Lyapunov exponent as a function of the rescaled inverse temperature $f\beta$ for fixed $f$. From the discussion above, as well as from \cref{fig:mss-convergence-temperature}, it is also clear that our system approaches the MSS bound for fixed $f$, if the temperature becomes sufficiently small. This means that we have a class of $f$-dimensional quantum systems with chaotic classical limits that become maximally fast scramblers in the low-temperature limit.  

Moreover, the simple form of \cref{eq:small-dim-lyapunov} suggests that upon rescaling by $1/f$, the quantum Lyapunov exponent should be described by a universal function of $f\beta$ with no (or only very weak) additional dependence on the dimension. The slightly opaque solid lines show $(1/f)\Lambda(f\beta)$, obtained from the full numerical solution of the stationarity condition \eqref{eq:stationarity} for two representative dimensions, $f\!=\!301$ and $f\!=\!4301$, in units where $\hbar=1,m=2,L=1$. The two numerical curves are so close to each other that their difference cannot be resolved in the plot. This, together with further analysis for other $f$ values, indeed points towards a unique curve $(1/f)\Lambda(f\beta)$ for describing the quantum Lyapunov exponent for large $f$. 

Furthermore the convergence towards the chaos bound at fixed $f$ suggests that even without taking the JT gravity limit, there might be  relatively simple gravitational duals.
This notion is supported by topological gravity, also known as the Airy model (see e.g. \cite{Kontsevich1992,saad_convergent_2022}). The density of states of this theory is identical to the one of our model in $f=3$ dimensions. At intermediate finite dimensions, our model can be seen as interpolating between topological and JT gravity by changing the dimension, in a manner similar, but not identical, to the $(2,2p-1)$ minimal string \cite{Saad2019,mertens_liouville_2021}, which has recently been found to admit a black-hole-like geometry \cite{Gregori2021}.

A somewhat similar behavior of the quantum Lyapunov exponent has been found in the SYK model in \cite{Kobrin2020a}. While for simple gravitational theories such as Einstein gravity, one expects $\Lambda$ to saturate the MSS bound, stringy corrections \cite{shenker_stringy_2015} can hinder the development of chaos and decrease the quantum Lyapunov exponent. This effect has also been discussed explicitly for the Schwarzian theory in \cite{qi_quantum_2019}, and a description in terms of so-called scramblon modes has been shown to be applicable in SYK-like models \cite{choi_effective_2023}. The submaximal chaos apparent in the quantum Lyapunov exponent in \cref{fig:mss-convergence-temperature}, viewed in this light, therefore hints at an interesting, as yet unexplored interpretation of the high-temperature regime of our model (at finite dimension) in terms of a more complicated gravitational theory with stringy (or similar) effects that disappear at low temperature, and that might even be explicable in terms of the corrections \eqref{eq:high-dim-rho} to the infinite-dimensional pure JT gravity density of states \eqref{eq:rho-infinity}.


\section{Subleading corrections}\label{sec:corrections}
As already mentioned above, the term whose growth rate is bounded by $2\pi/\hbar\beta$ in the out-of-time ordered correlator $F(t)$ is predicted \cite{Maldacena2016} to only be the first in an expansion,
\begin{equation}
    F_d-F(t)=\hbar^2 f_1e^{\frac{2\pi}{\hbar\beta}t}+\order{\hbar^4},\label{eq:otoc-approx}
\end{equation}
where $F_d$ is the factorized value of $F(t)$, as introduced in \cref{sec:otoc}. Indeed, defining moments
\begin{equation}
    \mu_J(t)=e^{\frac{4\pi J}{\hbar\beta}t}\int_{t-i\hbar\beta/4}^{t+i\hbar\beta/4}dt'e^{-\frac{2\pi}{\hbar\beta}\left(t'-i\hbar\beta/4\right)(2J+1)}(F(t')-F_d),
\end{equation}
the analytic structure of $F(t)$ imposes a set of conditions on the moments \cite{kundu_subleading_2022}:
\begin{align}
    0<\mu_J(t)&<\frac{2\hbar\beta F_d}{\pi(2J+1)}e^{-\frac{2\pi}{\hbar\beta}t},\label{eq:positivity}\\
    \mu_{J+1}(t)&<\mu_J(t),\label{eq:monotonicity}\\
    \mu_{J+1}(t)^2&\leq\mu_J(t)\mu_{J+2}(t).\label{eq:log-convexity}
\end{align}
These conditions must be satisfied by the out-of-time ordered correlator, and \cref{eq:positivity,eq:monotonicity} imply that at late enough times\footnote{A bit more precisely: the conditions \cref{eq:positivity,eq:monotonicity} imply that there have to be corrections to the maximal Lyapunov growth \eqref{eq:otoc-approx} of the form \eqref{eq:otoc-better-approx} with $\Lambda_2>\frac{2\pi}{\hbar\beta}$. These corrections then imply the existence of a timescale $t_1\ll t^*$ where the approximation \eqref{eq:otoc-approx} to the OTOC breaks down and the second term in \cref{eq:otoc-better-approx} starts to dominate \cite{kundu_subleading_2022}.}, corrections to the Lyapunov growth of e.g. the form 
\begin{equation}
    F_d-F(t)=\hbar^2f_1e^{\frac{2\pi}{\hbar\beta}t}+\hbar^4f_2e^{\Lambda_2t}+\order{\hbar^6}\label{eq:otoc-better-approx}
\end{equation}
must appear in systems with maximal Lyapunov growth $\Lambda=\frac{2\pi}{\hbar\beta}$. Further accounting for \cref{eq:log-convexity} produces a bound on this new, subleading exponential growth as well,
\begin{equation}
    \Lambda_2\leq\frac{6\pi}{\hbar\beta}.\label{eq:subleading-bound}
\end{equation}
The saturation of this equality again forces similar late time corrections $\Lambda_3,\Lambda_4,\ldots$, which are then again bounded by $\frac{10\pi}{\hbar\beta},\frac{14\pi}{\hbar\beta},\ldots$ and so on, by repeatedly applying \cref{eq:positivity,eq:monotonicity,eq:log-convexity} \cite{kundu_extremal_2022}. Since our system, as we have shown in \cref{sec:leading}, saturates the MSS bound at leading order in the OTOC, and the Wigner-Moyal quantization gives a systematic way to compute $\hbar$ corrections to the leading result, it is natural to examine those corrections and determine whether subleading bounds are saturated as well. In this section, we will attempt to characterize the first nonzero correction to the leading order OTOC in our system and estimate its growth rate. However, as we will see, the complexity of the computation is drastically higher than at leading order.

To recap some of the observations in \cref{sec:wigner-moyal}, we expect corrections to the leading order in $\hbar$ from
\begin{enumerate}
    \item the Weyl symbol for the time-evolved operator $P_t$ (though not for $X$),
    \item the Moyal bracket and the $\star$ product,
    \item and the Wigner function $W(x,p)$.
\end{enumerate}
Naively, one might also expect these corrections to be sensitive to the non-flat metric of the configuration space, since Weyl symbols in configuration spaces must be modified according to \cref{eq:weyl-symbol-def}, i.e. by including the determinant $g(x)$ of the configuration space metric at $x$. Given the essential singularity at $\hbar=0$ in the prescription \eqref{eq:weyl-symbol-def}, tracing the influence of the metric on the small $\hbar$ expansion of a Weyl symbol is a highly nontrivial endeavor. Fortunately, the formalism developed by Osborn and Molzahn \cite{osborn_moyal_1995} to determine Weyl symbols of Heisenberg operators takes care of the metric dependence automatically via the classical flow that enters the computation. Since the Weyl symbols of initial time Weyl-ordered polynomials in $X,P$ remain unchanged by the curved configuration space, the only explicit modification of the Weyl symbol computation we have to account for come from $\hbar$ corrections to the Hamiltonian, which we discuss in appendix \ref{app:weyl-symbol}. In the notation of \cref{sec:wigner-moyal}, our Hamiltonian's Weyl symbol has the form \cite{marinov_path_1980}
\begin{align}
    h_c&=\frac{1}{2m}p_ig^{ij}(x)p_j,\\
    h_2&=\frac{R}{12m},\\
    h_{r\neq2}&=0,
\end{align}
where $R$ is the Ricci scalar of the manifold. Due to the vanishing of $h_1$, we can conclude from \cref{eq:z1-deq} that
\begin{equation}
    z_1=0,
\end{equation}
and crucially, $h_2$ is a constant in the system we consider, leading to a simplification of \cref{eq:z2-deq}. We can therefore determine the first nonzero correction to the quantum trajectory $Z$ by integrating \cref{eq:z2-deq} according to \cref{eq:zr-integrated}, and find
\begin{equation}
    \begin{aligned}    z_2(t;\zeta)=\int_0^tds&\nabla\gamma(t,0\vert\zeta)J\nabla\gamma(s,0\vert\zeta)^TJ^{-1}\\
    &\times\left[-\frac{1}{8}\Pi^2(\gamma\cdot\nabla)^2+\frac{1}{12}\Pi_{12}\Pi_{23}(\gamma\cdot\nabla)^3\right]J\nabla h_c(t,\gamma(t,s\vert\zeta)).
    \end{aligned}\label{eq:quantum-trajectory-correction}
\end{equation}
We can simplify this a bit further in order to facilitate the discussion that follows, by realizing that
\begin{equation}
    \nabla\gamma(t,0\vert\zeta)\equiv M(t)
\end{equation}
is simply the monodromy of the classical flow. For a Hamiltonian flow, the monodromy preserves the symplectic form \cite{arnold_mathematical_1997},
\begin{equation}
    M(t)^{T}JM(t)=J,
\end{equation}
and hence, we can combine
\begin{equation}
    M(t)JM(s)^{T}J^{-1}=M(t)M(s)^{-1}=\nabla\gamma(t,s\vert\zeta),
\end{equation}
since $J^2=-1$ and multiplying the above equation with $M^{-1}$ from the right and $J^{-1}$ from the left. With this rewriting, we can express the correction to the quantum trajectory as
\begin{equation}
    \begin{aligned}    z_2(t;\zeta)=\int_0^tds&\nabla\gamma(t,s\vert\zeta)
    \left[-\frac{1}{8}\Pi^2(\gamma\cdot\nabla)^2+\frac{1}{12}\Pi_{12}\Pi_{23}(\gamma\cdot\nabla)^3\right]J\nabla h_c(t,\gamma(t,s\vert\zeta)).
    \end{aligned}\label{eq:quantum-trajectory-correction-2}
\end{equation}

To the desired order in $\hbar$, this accounts for all contributions to the OTOC coming from the Weyl symbols themselves. The remaining corrections stem from the Moyal bracket,
\begin{equation}
    \{\cdot,\cdot\}_M=\{\cdot,\cdot\}-\frac{\hbar^2}{12}\Pi^3(\cdot,\cdot)+\order{\hbar^4},
\end{equation}
and the $\star$ product,
\begin{equation}
    \star=\cdot+\frac{i\hbar}{2}\Pi-\frac{\hbar^2}{8}\Pi^2+\order{\hbar^3}.
\end{equation}
Note however, that in \cref{eq:otoc-weyl}, the $\star$ product is taken between two copies of the same object. Plugging in the $\hbar$ expansion for the Moyal brackets, the only terms appearing at $\order{\hbar}$ are the two ``fully classical'' ones, and the $\star$ product of these vanishes, cf. \cref{fn:asymmetry}.

There may be further corrections from the Wigner function which we have not considered. However, as we are interested only in the (exponential) growth of the OTOC, we argue for neglecting them, since they are generic for any correlation function of any set of operators, depending only on the state. For this reason, we do not expect them to contribute in an interesting manner to the growth rate.

All assembled then, the first correction to \cref{eq:otoc-weyl} is
\begin{equation}
        \hbar^4\Bigg(-\{p_2(t),x\}\{p_t,x\}
        +\frac{1}{6}\{p_t,x\}\Pi^3(p_t,x)
        +\frac{1}{8}\Pi^2\bigg(\{p_t,x\},\{p_t,x\}\bigg)\Bigg),
        \label{eq:correction}
\end{equation}
where we generalized the notation in \cref{eq:ansatz} to $W_{P_t}$, recalling that $p_0(t)=p_t$, and suppressed the dependence on the initial condition $\zeta$. At present, we do not know how to evaluate this expression or compute $\Lambda_2$ from it, although numerical estimates for some simple manifolds might be possible.

As a very rough estimate, we could use as a guideline that
\begin{equation}
    \{p_t,x\}\sim e^{\lambda t}.
\end{equation}
If we take this to mean that phase space derivatives acting on time-evolved quantities produce Lyapunov growth $e^{\lambda t}$, then we can give a broad-strokes prediction for the time dependence of each term in \cref{eq:correction}. Most obviously, for the middle term, we have the following:
\begin{equation}
    \Pi^3(p_t,x)=\sum_{k=0}^3(-1)^k\binom{3}{k}\left(\frac{\partial^k}{\partial p^k}\frac{\partial^{3-k}}{\partial x^{3-k}}p_t\right)\left(\frac{\partial^{3-k}}{\partial p^{3-k}}\frac{\partial^k}{\partial x^k}x\right).
\end{equation}
Now in this sum, if $k\neq 3$, the second term produces derivatives of initial $x$ wrt initial $p$, which vanish. But if $k=3$, we have $\partial^3x/\partial x^3=0$. Hence, this contribution vanishes identically,
\begin{equation}
    \{p_t,x\}\Pi^3(p_t,x)=0.
\end{equation}

For the third term, a superficial examination yields that each of the Poisson brackets acted upon by $\Pi^2$ already grows exponentially, so each application of the Poisson bivector should produce a factor $e^{2\lambda t}$, giving a total growth of $e^{6\lambda t}$. Indeed, we can evaluate this expression:
\begin{align}
    \Pi^2&(\{p_t,x\},\{p_t,x\})\\
    &=\sum_{k=0}^2(-1)^k\binom2{k}\left(\frac{\partial^k}{\partial p^k}\frac{\partial^{2-k}}{\partial x^{2-k}}\pdv{p_t}{p}\right)\left(\frac{\partial^{2-k}}{\partial p^{2-k}}\frac{\partial^{k}}{\partial x^{k}}\pdv{p_t}{p}\right)\\
    &=2\frac{\partial^3p_t}{\partial x^2\partial p}\frac{\partial^3p_t}{\partial p^3}-2\left(\frac{\partial^3 p_t}{\partial x\partial^2 p}\right)^2.
\end{align}
Evidently, this includes 6 total derivatives of the time-evolved momentum. A sensible expectation is then that the term grows no more strongly than
\begin{equation}
    \Pi^2(\{p_t,x\},\{p_t,x\})\sim e^{6\lambda t},\label{eq:pi2-term}
\end{equation}
as anticipated, and still in agreement with the bound \eqref{eq:subleading-bound}.

Finally, \cref{eq:quantum-trajectory-correction-2} is the hardest term to evaluate explicitly, since it depends in a nontrivial manner on the classical solutions on the hyperbolic manifold $\mathcal{M}$, which are not generally known analytically. The only way to provide any estimate at the moment is to be even more speculative. 

In the simpler free particle case on $\mathds{H}^f$, the $\gamma\cdot\nabla$ derivatives act in more or less the same way as the $J\nabla$ on the Hamiltonian, i.e. they are derivatives along the flow direction. The $\Pi$ derivatives are not directional and thus act differently. Naively counting all derivatives, the integrand might grow as fast as $8\lambda$. If we use the refined assumption that only the $\Pi$ derivatives produce exponential growth however, we can bound the growth of the integrand to at most $5\lambda$. Finally, when looking at the free particle solutions \cite{georgiou_space_2007}, the only function of the initial coordinates that we take $\Pi$ type derivatives of that seems to be sufficiently complicated to produce exponential growth are derivatives of the conformal factor of the metric:
\begin{equation}
    g_{ij}(x)=\Omega(x)\delta_{ij}, \quad\Omega(x)=\frac{2}{1-\abs{x}^2}.
\end{equation}
Since this only depends on $x$, it is conceivable that only the $x$ derivatives in $\Pi$ produce exponential growth, bounding the total exponent of the integral to $3\lambda$, although this is not the expected behavior in typical chaotic quantum systems.

If we then plug these estimates into the actual contribution to the OTOC, we estimate the growth to be bounded by
\begin{equation}
    -\{p_2(t),x\}\{p_t,x\}\sim\begin{cases}
        e^{10\lambda t}&\mathrm{if~all~derivatives~contribute,}\\
        e^{7\lambda t}&\mathrm{if~only~}\Pi\mathrm{~derivatives~contribute,}\\
        e^{5\lambda t}&\mathrm{if~only~}x~\mathrm{~derivatives~in~}\Pi\mathrm{~contribute.}
    \end{cases}\label{eq:first-term}
\end{equation}
In our particular system, we were able to find in \cref{sec:leading} that in the limit $f\to\infty$, the classical Lyapunov exponent is evaluated at the saddle point as $\frac{\pi}{\hbar\beta}$, which would mean that the total growth of the $\hbar^4$ correction to the OTOC grows with
\begin{equation}
    \Lambda_2=\begin{cases}
        \frac{10\pi}{\hbar\beta}&\mathrm{if~all~derivatives~contribute,}\\
        \frac{7\pi}{\hbar\beta}&\mathrm{if~only~}\Pi\mathrm{~derivatives~contribute,}\\
        \frac{6\pi}{\hbar\beta}&\mathrm{if~only~}x~\mathrm{~derivatives~in~}\Pi\mathrm{~contribute,}
    \end{cases},\label{eq:Lambda-2}
\end{equation}
since in the third case, the contribution from \cref{eq:pi2-term} dominates the one from \cref{eq:first-term}. This means that only in the third case, we could guarantee the bound \eqref{eq:subleading-bound} being respected at all, and our estimate would leave room for the bound to be saturated as well. Interestingly, this saturation would, if at all, arise from the correction to the $\star$ product, not the quantum trajectory. In the other cases meanwhile, we can not say anything definite, but it is at least encouraging that our method gives an estimate for the growth bound that is reasonably close to \cref{eq:subleading-bound}, and not orders of magnitude off.

We can compare these estimates to the Schwarzian theory, which describes both JT gravity and the SYK model. In this theory, the out-of-time ordered correlator can be computed exactly in the limit $\beta\ll1$, and is found to be (in the usual natural units for JT gravity) \cite{maldacena_conformal_2016,kundu_extremal_2022}
\begin{equation}
    \frac{F(t)}{F_d}=\frac{1}{z^{2\Delta}}U(2\Delta,1,1/z), \qquad z=\frac{\beta}{8\pi}e^{\frac{2\pi t}{\beta}},\label{eq:otoc-jt}
\end{equation}
for a pair of operators $V,W$ with scaling dimension $\Delta$. Here, $U(a,b,y)$ is the confluent hypergeometric function. At early enough times, it has an asymptotic expansion in terms of the generalized hypergeometric function \cite{andrews_special_1999}
\begin{equation}
    U(a,b,y)\sim y^{-a}\ _2F_0(a,a-b+1;;-1/y),
\end{equation}
where
\begin{equation}
    _2F_0(a_1,a_2;;z)=\frac{1}{\Gamma(a_1)\Gamma(a_2)}\sum_{n=0}^{\infty}\Gamma(n+a_1)\Gamma(n+a_2)\frac{z^n}{n!}.
\end{equation}
Expanding \cref{eq:otoc-jt} up to second order in small $z$, we find
\begin{equation}
    \frac{F(t)}{F_d}=1-\frac{\beta\Delta^2}{2\pi}e^{\frac{2\pi t}{\beta}}+\frac{\beta^2\Delta^2(2\Delta+1)^2}{32\pi^2}e^{\frac{4\pi t}{\beta}}+\ldots\label{eq:otoc-jt-expansion}
\end{equation}
Clearly, the subleading correction grows with $\Lambda_2=\frac{4\pi}{\beta}$, i.e. far away from the bound \eqref{eq:subleading-bound}. This tension with our rough estimate \eqref{eq:Lambda-2} suggests that a more careful examination of the OTOC in our system is needed. It should be noted however, that \cref{eq:otoc-jt-expansion} is an expansion for small $\beta$, whereas our system's OTOC at fixed dimension approaches the MSS bound only for relatively large $\beta$, cf. \cref{fig:mss-convergence-temperature}, meaning that the two behaviours need not necessarily agree.
\section{Summary}
In this work, we have developed a method to compute the canonical quantum Lyapunov exponent in systems with a large number of degrees of freedom, using Wigner-Moyal phase space quantization as a tool. Since the leading order in $\hbar$ in this formalism is simply equivalent to the classical Poisson algebra on the phase space, the calculation requires only two inputs: the classical microcanonical Lyapunov exponent, and the density of states. Remarkably, for a particle sliding on a high-dimensional hyperbolic manifold, the system that was shown in \cite{haneder_beyond_2025} to reproduce correlation functions of JT gravity in the limit of the configuration space dimension going to infinity, both of these quantities, including all quantum corrections to the density of states, are known in the literature, facilitating the computation of the quantum Lyapunov exponent via the solution of a simple saddle point condition \eqref{eq:stationarity}.

We have shown that at large enough dimension, the quantum Lyapunov exponent interpolates between the maximal value provided by the Maldacena-Shenker-Stanford bound at low temperatures, and a classical regime described by the equipartition theorem at high temperatures, cf. \cref{fig:mss-convergence-temperature}. In the limit of the dimension $f\to\infty$, the Lyapunov exponent saturates the MSS bound for all $\beta$, as evident from \cref{fig:mss-convergence-dimension} and \cref{eq:infinite-lyapunov}, adding to the evidence in \cite{haneder_beyond_2025} of the system being dual to JT gravity in the $f\to\infty$ limit, and indeed showing gravitational signatures even at finite dimension and low temperatures. This latter point in particular opens up the interesting possibility of studying large-but-finite dimensional Hadamard-Gutzwiller like models as potential duals of gravitational systems in the spirit of the $(2,p)$ minimal string, with a possible interpretation of the difference between the infinite- and finite-dimensional spectral densities in terms of stringy corrections \cite{shenker_stringy_2015,qi_quantum_2019,choi_effective_2023}. 

We were also able to find a fairly compact expression, \cref{eq:correction}, for the first subleading correction to the Weyl symbol of the OTOC using the formalism developed in \cite{osborn_moyal_1995}, as well as partially estimate its growth exponent \eqref{eq:Lambda-2}. While we were not able to verify that the bound of \cite{kundu_subleading_2022,kundu_extremal_2022} is observed, we could nevertheless restrict the growth of the first correction to the OTOC to be at least not much larger than allowed by \cref{eq:subleading-bound}. A more thorough evaluation is complicated by the need to determine the solutions of the equations of motion on a high-dimensional, compact manifold, as well as quantum corrections to the Weyl symbols of the Heisenberg operators $X_t,P_t$. The limit is also fairly discontinuous, requiring in principle the solution of new equations of motion every time $f$ is increased. Given the difficulty of finding a reliable estimate for the growth rate of the subleading correction in \cref{sec:corrections}, a more careful investigation using e.g. the easily generalizable solutions of the free motion on the hyperbolic $f$-space, or numerical solutions on a fixed high dimension, might provide valuable insights, especially in light of the possibility of extremal chaoticity that we were not able to exclude, and which would be unexpected in a system dual to JT gravity.

\acknowledgments
We thank Maximilian Kieler for valuable comments on the interpretation of higher-order phase space derivatives and Torsten Weber, Mathias Steinhuber and Georg Maier for useful discussions. We acknowledge financial support from the Deutsche Forschungsgemeinschaft (German Research Foundation) through Ri681/15-1 (project number 456449460) within the Reinhart-Koselleck Programme.
\bibliography{references}

@article{Maldacena2016,
    title = {A bound on chaos},
    volume = {2016},
    issn = {1029-8479},
    url = {http://link.springer.com/10.1007/JHEP08(2016)106},
    doi = {10.1007/JHEP08(2016)106},
    number = {8},
    urldate = {2019-10-16},
    journal = {Journal of High Energy Physics},
    author = {Maldacena, Juan and Shenker, Stephen H. and Stanford, Douglas},
    month = aug,
    year = {2016},
    note = {arXiv: 1503.01409
Publisher: Springer Verlag},
    pages = {106},
}

@article{garcia-mata_out--time-order_2023,
    title = {Out-of-time-order correlations and quantum chaos},
    volume = {18},
    issn = {1941-6016},
    url = {http://www.scholarpedia.org/article/Out-of-time-order_correlations_and_quantum_chaos},
    doi = {10.4249/scholarpedia.55237},
    language = {en},
    number = {4},
    urldate = {2025-06-24},
    journal = {Scholarpedia},
    author = {Garc\'{i}a-Mata, Ignacio and Jalabert, Rodolfo A. and Wisniacki, Diego Ariel},
    month = apr,
    year = {2023},
    pages = {55237},
}

@article{kundu_subleading_2022,
    title = {Subleading {Bounds} on {Chaos}},
    volume = {2022},
    issn = {1029-8479},
    url = {http://arxiv.org/abs/2109.03826},
    doi = {10.1007/JHEP04(2022)010},
    number = {4},
    urldate = {2025-04-29},
    journal = {Journal of High Energy Physics},
    author = {Kundu, Sandipan},
    month = apr,
    year = {2022},
    note = {arXiv:2109.03826 [hep-th]},
    pages = {10},
}

@article{kundu_extremal_2022,
    title = {Extremal {Chaos}},
    volume = {2022},
    issn = {1029-8479},
    url = {http://arxiv.org/abs/2109.08693},
    doi = {10.1007/JHEP01(2022)163},
    number = {1},
    urldate = {2025-04-29},
    journal = {Journal of High Energy Physics},
    author = {Kundu, Sandipan},
    month = jan,
    year = {2022},
    note = {arXiv:2109.08693 [hep-th]},
    pages = {163},
}

@article{gneiting_quantum_2013,
    title = {Quantum phase-space representation for curved configuration spaces},
    volume = {88},
    url = {https://link.aps.org/doi/10.1103/PhysRevA.88.062117},
    doi = {10.1103/PhysRevA.88.062117},
    number = {6},
    urldate = {2025-02-04},
    journal = {Physical Review A},
    author = {Gneiting, Clemens and Fischer, Timo and Hornberger, Klaus},
    month = dec,
    year = {2013},
    pages = {062117},
}

@article{osborn_moyal_1995,
    title = {Moyal {Quantum} {Mechanics}: {The} {Semiclassical} {Heisenberg} {Dynamics}},
    volume = {241},
    issn = {0003-4916},
    shorttitle = {Moyal {Quantum} {Mechanics}},
    url = {https://www.sciencedirect.com/science/article/pii/S0003491685710573},
    doi = {10.1006/aphy.1995.1057},
    number = {1},
    urldate = {2025-04-29},
    journal = {Annals of Physics},
    author = {Osborn, T. A. and Molzahn, F. H.},
    month = jul,
    year = {1995},
    pages = {79--127},
}

@article{dewitt_point_1952,
    title = {Point {Transformations} in {Quantum} {Mechanics}},
    volume = {85},
    url = {https://link.aps.org/doi/10.1103/PhysRev.85.653},
    doi = {10.1103/PhysRev.85.653},
    number = {4},
    urldate = {2025-07-22},
    journal = {Physical Review},
    author = {DeWitt, Bryce Seligman},
    month = feb,
    year = {1952},
    pages = {653--661},
}

@article{dewitt_dynamical_1957,
    title = {Dynamical {Theory} in {Curved} {Spaces}. {I}. {A} {Review} of the {Classical} and {Quantum} {Action} {Principles}},
    volume = {29},
    url = {https://link.aps.org/doi/10.1103/RevModPhys.29.377},
    doi = {10.1103/RevModPhys.29.377},
    number = {3},
    urldate = {2025-07-22},
    journal = {Reviews of Modern Physics},
    author = {DeWitt, Bryce S.},
    month = jul,
    year = {1957},
    pages = {377--397},
}

@article{marinov_path_1980,
    title = {Path integrals in quantum theory: {An} outlook of basic concepts},
    volume = {60},
    issn = {0370-1573},
    shorttitle = {Path integrals in quantum theory},
    url = {https://www.sciencedirect.com/science/article/pii/0370157380901118},
    doi = {10.1016/0370-1573(80)90111-8},
    number = {1},
    urldate = {2025-04-29},
    journal = {Physics Reports},
    author = {Marinov, M. S.},
    month = apr,
    year = {1980},
    pages = {1--57},
}

@book{gaspard_chaos_1998,
    address = {Cambridge},
    series = {Cambridge {Nonlinear} {Science} {Series}},
    title = {Chaos, {Scattering} and {Statistical} {Mechanics}},
    isbn = {9780521395113},
    url = {https://www.cambridge.org/core/books/chaos-scattering-and-statistical-mechanics/B2F337764BE796C22EE7272DF78DF7F6},
    urldate = {2025-07-24},
    publisher = {Cambridge University Press},
    author = {Gaspard, Pierre},
    year = {1998},
    doi = {10.1017/CBO9780511628856},
}

@article{groenewold_principles_1946,
    title = {On the principles of elementary quantum mechanics},
    volume = {12},
    issn = {0031-8914},
    url = {https://www.sciencedirect.com/science/article/pii/S0031891446800594},
    doi = {10.1016/S0031-8914(46)80059-4},
    number = {7},
    urldate = {2025-07-29},
    journal = {Physica},
    author = {Groenewold, H. J.},
    month = oct,
    year = {1946},
    pages = {405--460},
}

@article{moyal_quantum_1949,
    title = {Quantum mechanics as a statistical theory},
    volume = {45},
    issn = {1469-8064, 0305-0041},
    url = {https://www.cambridge.org/core/journals/mathematical-proceedings-of-the-cambridge-philosophical-society/article/quantum-mechanics-as-a-statistical-theory/9D0DC7453AD14DB641CF8D477B3C72A2},
    doi = {10.1017/S0305004100000487},
    language = {en},
    number = {1},
    urldate = {2025-07-29},
    journal = {Mathematical Proceedings of the Cambridge Philosophical Society},
    author = {Moyal, J. E.},
    month = jan,
    year = {1949},
    pages = {99--124},
}

@article{kontsevich_deformation_2003,
    title = {Deformation {Quantization} of {Poisson} {Manifolds}},
    volume = {66},
    issn = {1573-0530},
    url = {https://doi.org/10.1023/B:MATH.0000027508.00421.bf},
    doi = {10.1023/B:MATH.0000027508.00421.bf},
    language = {en},
    number = {3},
    urldate = {2025-07-29},
    journal = {Letters in Mathematical Physics},
    author = {Kontsevich, Maxim},
    month = dec,
    year = {2003},
    pages = {157--216},
}

@article{turiaci_inelastic_2019,
    title = {An inelastic bound on chaos},
    volume = {2019},
    issn = {1029-8479},
    url = {https://doi.org/10.1007/JHEP07(2019)099},
    doi = {10.1007/JHEP07(2019)099},
    language = {en},
    number = {7},
    urldate = {2025-08-04},
    journal = {Journal of High Energy Physics},
    author = {Turiaci, Gustavo J.},
    month = jul,
    year = {2019},
    pages = {99},
}

@article{jahnke_chaos_2019,
    title = {On the chaos bound in rotating black holes},
    volume = {2019},
    issn = {1029-8479},
    url = {https://doi.org/10.1007/JHEP05(2019)037},
    doi = {10.1007/JHEP05(2019)037},
    language = {en},
    number = {5},
    urldate = {2025-08-04},
    journal = {Journal of High Energy Physics},
    author = {Jahnke, Viktor and Kim, Keun-Young and Yoon, Junggi},
    month = may,
    year = {2019},
    pages = {37},
}

@article{larkin_quasiclassical_1969,
    title = {Quasiclassical {Method} in the {Theory} of {Superconductivity}},
    volume = {28},
    journal = {Soviet Physics JETP},
    author = {Larkin, A. I. and Ovchinnikov, Yu N.},
    year = {1969},
    pages = {1200--1205},
}

@article{Kobrin2020a,
    title = {Many-{Body} {Chaos} in the {Sachdev}-{Ye}-{Kitaev} {Model}},
    volume = {126},
    issn = {0031-9007},
    url = {http://arxiv.org/abs/2002.05725},
    doi = {10.1103/PhysRevLett.126.030602},
    number = {3},
    urldate = {2021-03-24},
    journal = {Physical Review Letters},
    author = {Kobrin, Bryce and Yang, Zhenbin and Kahanamoku-Meyer, Gregory D. and Olund, Christopher T. and Moore, Joel E. and Stanford, Douglas and Yao, Norman Y.},
    month = jan,
    year = {2021},
    note = {arXiv: 2002.05725
Publisher: American Physical Society},
    pages = {030602},
}

@article{shenker_multiple_2014,
    title = {Multiple shocks},
    volume = {2014},
    issn = {1029-8479},
    url = {https://doi.org/10.1007/JHEP12(2014)046},
    doi = {10.1007/JHEP12(2014)046},
    language = {en},
    number = {12},
    urldate = {2025-08-04},
    journal = {Journal of High Energy Physics},
    author = {Shenker, Stephen H. and Stanford, Douglas},
    month = dec,
    year = {2014},
    pages = {46},
}

@article{Gregori2021,
    title = {From {Minimal} {Strings} towards {Jackiw}-{Teitelboim} {Gravity}: {On} their {Resurgence}, {Resonance}, and {Black} {Holes}},
    url = {http://arxiv.org/abs/2108.11409},
    archivePrefix = {arXiv},
    arxivID = {2108.11409},
    eprint = {2108.11409},
    urldate = {2021-08-27},
    url = {https://arxiv.org/abs/2108.11409},
    author = {Gregori, Paolo and Schiappa, Ricardo},
    month = aug,
    year = {2021},
    note = {arXiv: 2108.11409},
}

@article{aurich_periodic_1988,
    title = {On the periodic orbits of a strongly chaotic system},
    volume = {32},
    issn = {01672789},
    url = {https://linkinghub.elsevier.com/retrieve/pii/0167278988900681},
    doi = {10.1016/0167-2789(88)90068-1},
    language = {en},
    number = {3},
    urldate = {2023-05-23},
    journal = {Physica D: Nonlinear Phenomena},
    author = {Aurich, R. and Steiner, F.},
    month = dec,
    year = {1988},
    pages = {451--460},
}

@article{Aurich1989,
    title = {Periodic-orbit sum rules for the {Hadamard}-{Gutzwiller} model},
    volume = {39},
    issn = {01672789},
    url = {https://linkinghub.elsevier.com/retrieve/pii/0167278989900031},
    doi = {10.1016/0167-2789(89)90003-1},
    number = {2-3},
    urldate = {2021-02-19},
    journal = {Physica D: Nonlinear Phenomena},
    author = {Aurich, R. and Steiner, F.},
    month = oct,
    year = {1989},
    pages = {169--193},
}

@article{poincare_sur_1908,
    title = {Sur l'uniformisation des fonctions analytiques},
    volume = {31},
    issn = {0001-5962, 1871-2509},
    url = {https://projecteuclid.org/journals/acta-mathematica/volume-31/issue-none/Sur-luniformisation-des-fonctions-analytiques/10.1007/BF02415442.full},
    doi = {10.1007/BF02415442},
    number = {none},
    urldate = {2025-08-06},
    journal = {Acta Mathematica},
    author = {Poincar\'{e}, H.},
    month = jan,
    year = {1908},
    pages = {1--63},
}

@article{selberg_harmonic_1956,
    title = {Harmonic analysis and discontinuous groups in weakly symmetric {Riemannian} spaces with applications to {Dirichlet} series},
    volume = {20},
    issn = {0019-5839},
    url = {https://mathscinet.ams.org/mathscinet-getitem?mr=88511},
    urldate = {2023-03-28},
    journal = {The Journal of the Indian Mathematical Society. New Series},
    author = {Selberg, A.},
    year = {1956},
    mrnumber = {88511},
    pages = {47--87},
}

@article{Hannay1984,
    title = {Periodic orbits and a correlation function for the semiclassical density of states},
    volume = {17},
    issn = {0305-4470},
    url = {https://iopscience.iop.org/article/10.1088/0305-4470/17/18/013},
    doi = {10.1088/0305-4470/17/18/013},
    number = {18},
    urldate = {2020-11-05},
    journal = {Journal of Physics A: Mathematical and General},
    author = {Hannay, J. H. and Ozorio De Almeida, A M},
    month = dec,
    year = {1984},
    note = {Publisher: IOP Publishing},
    pages = {3429--3440},
}

@article{haneder_beyond_2025,
    title = {Beyond the ensemble paradigm in low-dimensional quantum gravity: {Schwarzian} density, quantum chaos, and wormhole contributions},
    volume = {111},
    shorttitle = {Beyond the ensemble paradigm in low-dimensional quantum gravity},
    url = {https://link.aps.org/doi/10.1103/rsrq-l6z8},
    doi = {10.1103/rsrq-l6z8},
    number = {12},
    urldate = {2025-08-07},
    journal = {Physical Review D},
    author = {Haneder, Fabian and Urbina, Juan Diego and Moreno, Camilo and Weber, Torsten and Richter, Klaus},
    month = jun,
    year = {2025},
    pages = {126015},
}

@article{jalabert_semiclassical_2018,
    title = {Semiclassical theory of out-of-time-order correlators for low-dimensional classically chaotic systems},
    volume = {98},
    url = {https://link.aps.org/doi/10.1103/PhysRevE.98.062218},
    doi = {10.1103/PhysRevE.98.062218},
    number = {6},
    urldate = {2025-08-07},
    journal = {Physical Review E},
    author = {Jalabert, Rodolfo A. and Garc\'{i}a-Mata, Ignacio and Wisniacki, Diego A.},
    month = dec,
    year = {2018},
    pages = {062218},
}

@incollection{randol_selberg_1984,
    title = {The {Selberg} {Trace} {Formula}},
    volume = {115},
    copyright = {https://www.elsevier.com/tdm/userlicense/1.0/},
    isbn = {978-0-12-170640-1},
    url = {https://linkinghub.elsevier.com/retrieve/pii/S0079816908608193},
    language = {en},
    urldate = {2024-06-05},
    booktitle = {Pure and {Applied} {Mathematics}},
    publisher = {Elsevier},
    author = {Randol, Burton},
    year = {1984},
    doi = {10.1016/S0079-8169(08)60819-3},
    pages = {266--302},
}

@article{bytsenko_quantum_1996,
    title = {Quantum fields and extended objects in space-times with constant curvature spatial section},
    volume = {266},
    issn = {0370-1573},
    url = {https://www.sciencedirect.com/science/article/pii/0370157395000534},
    doi = {10.1016/0370-1573(95)00053-4},
    number = {1},
    urldate = {2025-08-06},
    journal = {Physics Reports},
    author = {Bytsenko, Andrei A and Cognola, Guido and Vanzo, Luciano and Zerbini, Sergio},
    month = feb,
    year = {1996},
    pages = {1--126},
}

@article{david_chaos_2019,
    title = {Chaos bound in {Bershadsky}-{Polyakov} theory},
    volume = {2019},
    issn = {1029-8479},
    url = {https://doi.org/10.1007/JHEP10(2019)077},
    doi = {10.1007/JHEP10(2019)077},
    language = {en},
    number = {10},
    urldate = {2025-08-08},
    journal = {Journal of High Energy Physics},
    author = {David, Justin R. and Hollowood, Timothy J. and Khetrapal, Surbhi and Kumar, S. Prem},
    month = oct,
    year = {2019},
    pages = {77},
}

@article{perlmutter_bounding_2016,
    title = {Bounding the space of holographic {CFTs} with chaos},
    volume = {2016},
    issn = {1029-8479},
    url = {https://doi.org/10.1007/JHEP10(2016)069},
    doi = {10.1007/JHEP10(2016)069},
    language = {en},
    number = {10},
    urldate = {2025-08-08},
    journal = {Journal of High Energy Physics},
    author = {Perlmutter, Eric},
    month = oct,
    year = {2016},
    pages = {69},
}

@article{ullmo_orbital_1995,
    title = {Orbital {Magnetism} in {Ensembles} of {Ballistic} {Billiards}},
    volume = {74},
    url = {https://link.aps.org/doi/10.1103/PhysRevLett.74.383},
    doi = {10.1103/PhysRevLett.74.383},
    number = {3},
    urldate = {2025-08-08},
    journal = {Physical Review Letters},
    author = {Ullmo, Denis and Richter, Klaus and Jalabert, Rodolfo A.},
    month = jan,
    year = {1995},
    pages = {383--386},
}

@article{Kontsevich1992,
    title = {Mathematical {Physics} {Intersection} {Theory} on the {Moduli} {Space} of {Curves} and the {Matrix} {Airy} {Function}*},
    volume = {147},
    urldate = {2021-10-26},
    journal = {Commun. Math. Phys},
    author = {Kontsevich, Maxim},
    year = {1992},
    pages = {1--23},
}

@article{saad_convergent_2022,
    title = {A convergent genus expansion for the plateau},
    url = {http://arxiv.org/abs/2210.11565},
    urldate = {2022-11-03},
    publisher = {arXiv},
    archivePrefix = {arXiv},
    arxivID = {2210.11565},
    eprint = {2210.11565},
    author = {Saad, Phil and Stanford, Douglas and Yang, Zhenbin and Yao, Shunyu},
    month = oct,
    year = {2022},
    note = {arXiv:2210.11565 [hep-th, physics:nlin, physics:quant-ph]},
}

@article{Saad2019,
    title = {{JT} gravity as a matrix integral},
    url = {http://arxiv.org/abs/1903.11115},
    author = {Saad, Phil and Shenker, Stephen H. and Stanford, Douglas},
    archivePrefix = {arXiv},
    arxivID = {1903.11115},
    eprint = {1903.11115},
    month = mar,
    year = {2019},
    note = {arXiv: 1903.11115},
}

@article{mertens_liouville_2021,
    title = {Liouville quantum gravity -- holography, {JT} and matrices},
    volume = {2021},
    issn = {1029-8479},
    url = {https://doi.org/10.1007/JHEP01(2021)073},
    doi = {10.1007/JHEP01(2021)073},
    language = {en},
    number = {1},
    urldate = {2025-09-25},
    journal = {Journal of High Energy Physics},
    author = {Mertens, Thomas G. and Turiaci, Gustavo J.},
    month = jan,
    year = {2021},
    pages = {73},
}

@article{Sekino2008,
    title = {Fast scramblers},
    volume = {2008},
    issn = {1029-8479},
    url = {http://stacks.iop.org/1126-6708/2008/i=10/a=065?key=crossref.0eebed9ef6bfd908b2747d054e345a56},
    doi = {10.1088/1126-6708/2008/10/065},
    number = {10},
    urldate = {2019-10-16},
    journal = {Journal of High Energy Physics},
    author = {Sekino, Yasuhiro and Susskind, L.},
    month = oct,
    year = {2008},
    note = {arXiv: 0808.2096},
    pages = {065--065},
}

@article{yoshida_efficient_2017,
    title = {Efficient decoding for the {Hayden}-{Preskill} protocol},
    url = {http://arxiv.org/abs/1710.03363},
    doi = {10.48550/arXiv.1710.03363},
    urldate = {2025-09-29},
    publisher = {arXiv},
    archivePrefix = {arXiv},
    arxivID = {1710.03363},
    eprint = {1710.03363},
    author = {Yoshida, Beni and Kitaev, Alexei},
    month = oct,
    year = {2017},
    note = {arXiv:1710.03363 [hep-th]},
}

@article{gao_traversable_2021,
    title = {A traversable wormhole teleportation protocol in the {SYK} model},
    volume = {2021},
    issn = {1029-8479},
    url = {https://doi.org/10.1007/JHEP07(2021)097},
    doi = {10.1007/JHEP07(2021)097},
    language = {en},
    number = {7},
    urldate = {2025-09-25},
    journal = {Journal of High Energy Physics},
    author = {Gao, Ping and Jafferis, Daniel Louis},
    month = jul,
    year = {2021},
    pages = {97},
}

@article{jafferis_traversable_2022,
    title = {Traversable wormhole dynamics on a quantum processor},
    volume = {612},
    copyright = {2022 The Author(s), under exclusive licence to Springer Nature Limited},
    issn = {1476-4687},
    url = {https://www.nature.com/articles/s41586-022-05424-3},
    doi = {10.1038/s41586-022-05424-3},
    language = {en},
    number = {7938},
    urldate = {2025-09-25},
    journal = {Nature},
    author = {Jafferis, Daniel and Zlokapa, Alexander and Lykken, Joseph D. and Kolchmeyer, David K. and Davis, Samantha I. and Lauk, Nikolai and Neven, Hartmut and Spiropulu, Maria},
    month = dec,
    year = {2022},
    pages = {51--55},
}

@article{kobrin_experiments_2025,
    title = {Experiments implementing small commuting models lack gravitational features},
    volume = {643},
    copyright = {2025 The Author(s), under exclusive licence to Springer Nature Limited},
    issn = {1476-4687},
    url = {https://www.nature.com/articles/s41586-025-08939-7},
    doi = {10.1038/s41586-025-08939-7},
    language = {en},
    number = {8073},
    urldate = {2025-09-25},
    journal = {Nature},
    author = {Kobrin, Bryce and Schuster, Thomas and Yao, Norman Y.},
    month = jul,
    year = {2025},
    pages = {E17--E19},
}

@article{hayden_black_2007,
    title = {Black holes as mirrors: quantum information in random subsystems},
    volume = {2007},
    issn = {1126-6708},
    shorttitle = {Black holes as mirrors},
    url = {https://doi.org/10.1088/1126-6708/2007/09/120},
    doi = {10.1088/1126-6708/2007/09/120},
    language = {en},
    number = {09},
    urldate = {2025-09-30},
    journal = {Journal of High Energy Physics},
    author = {Hayden, Patrick and Preskill, John},
    month = sep,
    year = {2007},
    pages = {120},
}

@article{Shenker2014a,
    title = {Black holes and the butterfly effect},
    volume = {2014},
    issn = {10298479},
    url = {https://link.springer.com/article/10.1007/JHEP03(2014)067},
    doi = {10.1007/JHEP03(2014)067},
    number = {3},
    urldate = {2021-03-30},
    journal = {Journal of High Energy Physics},
    author = {Shenker, Stephen H. and Stanford, Douglas},
    month = mar,
    year = {2014},
    note = {arXiv: 1306.0622
Publisher: Springer Verlag},
    pages = {67},
}

@article{touil_information_2024,
    title = {Information scrambling -- {A} quantum thermodynamic perspective},
    volume = {146},
    issn = {0295-5075},
    url = {https://doi.org/10.1209/0295-5075/ad4413},
    doi = {10.1209/0295-5075/ad4413},
    language = {en},
    number = {4},
    urldate = {2025-09-29},
    journal = {Europhysics Letters},
    author = {Touil, Akram and Deffner, Sebastian},
    month = jun,
    year = {2024},
    pages = {48001},
}

@book{Haake2010a,
    address = {Berlin [u.a.]},
    title = {Quantum signatures of chaos},
    isbn = {978-3-642-05427-3},
    urldate = {2021-03-24},
    publisher = {Springer},
    author = {Haake, Fritz},
    year = {2010},
    note = {Series Title: Springer series in synergetics},
}

@article{wootters_single_1982,
    title = {A single quantum cannot be cloned},
    volume = {299},
    copyright = {1982 Springer Nature Limited},
    issn = {1476-4687},
    url = {https://www.nature.com/articles/299802a0},
    doi = {10.1038/299802a0},
    language = {en},
    number = {5886},
    urldate = {2025-09-30},
    journal = {Nature},
    author = {Wootters, W. K. and Zurek, W. H.},
    month = oct,
    year = {1982},
    pages = {802--803},
}

@article{dieks_communication_1982,
    title = {Communication by {EPR} devices},
    volume = {92},
    issn = {0375-9601},
    url = {https://www.sciencedirect.com/science/article/pii/0375960182900846},
    doi = {10.1016/0375-9601(82)90084-6},
    number = {6},
    urldate = {2025-09-30},
    journal = {Physics Letters A},
    author = {Dieks, D.},
    month = nov,
    year = {1982},
    pages = {271--272},
}

@article{sachdev_gapless_1993,
    title = {Gapless spin-fluid ground state in a random quantum {Heisenberg} magnet},
    volume = {70},
    url = {https://link.aps.org/doi/10.1103/PhysRevLett.70.3339},
    doi = {10.1103/PhysRevLett.70.3339},
    number = {21},
    urldate = {2025-10-01},
    journal = {Physical Review Letters},
    author = {Sachdev, Subir and Ye, Jinwu},
    month = may,
    year = {1993},
    pages = {3339--3342},
}

@misc{noauthor_alexei_nodate,
    title = {Alexei {Kitaev}, {Caltech}, {A} simple model of quantum holography (part 2)},
    url = {https://online.kitp.ucsb.edu/online/entangled15/kitaev2/},
    urldate = {2025-10-01},
    author = {Kitaev, Alexei},
}

@misc{kitaev_alexei_nodate,
    title = {Alexei {Kitaev}, {Caltech} \& {KITP}, {A} simple model of quantum holography (part 1)},
    url = {https://online.kitp.ucsb.edu/online/entangled15/kitaev/},
    urldate = {2025-10-01},
    author = {Kitaev, Alexei},
}

@article{Jackiw1985,
    title = {Lower dimensional gravity},
    volume = {252},
    issn = {0550-3213},
    doi = {10.1016/0550-3213(85)90448-1},
    number = {C},
    urldate = {2022-06-30},
    journal = {Nuclear Physics B},
    author = {Jackiw, Roman},
    month = jan,
    year = {1985},
    note = {Publisher: North-Holland},
    pages = {343--356},
}

@article{Teitelboim1983,
    title = {Gravitation and hamiltonian structure in two spacetime dimensions},
    volume = {126},
    issn = {0370-2693},
    doi = {10.1016/0370-2693(83)90012-6},
    number = {1-2},
    urldate = {2022-06-30},
    journal = {Physics Letters B},
    author = {Teitelboim, Claudio},
    month = jun,
    year = {1983},
    note = {Publisher: North-Holland},
    pages = {41--45},
}

@article{gutzwiller_periodic_1971,
    title = {Periodic {Orbits} and {Classical} {Quantization} {Conditions}},
    volume = {12},
    issn = {0022-2488},
    url = {https://doi.org/10.1063/1.1665596},
    doi = {10.1063/1.1665596},
    number = {3},
    urldate = {2025-10-02},
    journal = {Journal of Mathematical Physics},
    author = {Gutzwiller, Martin C.},
    month = mar,
    year = {1971},
    pages = {343--358},
}

@article{maldacena_conformal_2016,
    title = {Conformal symmetry and its breaking in two-dimensional nearly anti-de {Sitter} space},
    volume = {2016},
    issn = {2050-3911},
    url = {https://doi.org/10.1093/ptep/ptw124},
    doi = {10.1093/ptep/ptw124},
    number = {12},
    urldate = {2025-10-07},
    journal = {Progress of Theoretical and Experimental Physics},
    author = {Maldacena, Juan and Stanford, Douglas and Yang, Zhenbin},
    month = dec,
    year = {2016},
    pages = {12C104},
}

@book{andrews_special_1999,
    title = {Special {Functions}},
    isbn = {9780521789882},
    address = {Cambridge, United Kingdom},
    language = {en},
    publisher = {Cambridge University Press},
    author = {Andrews, George E. and Askey, Richard and Roy, Ranjan},
    year = {1999},
    note = {Google-Books-ID: kGshpCa3eYwC},
}

@article{georgiou_space_2007,
    title = {On the space of oriented geodesics of hyperbolic 3-space},
    url = {http://arxiv.org/abs/math/0702276},
    doi = {10.48550/arXiv.math/0702276},
    urldate = {2025-04-29},
    publisher = {arXiv},
    archivePrefix = {arXiv},
    arxivID = {math/0702276},
    eprint = {math/0702276},
    author = {Georgiou, Nikos and Guilfoyle, Brendan},
    month = mar,
    year = {2007},
    note = {arXiv:math/0702276},
}

@article{hashimoto_bound_2022,
    title = {Bound on energy dependence of chaos},
    volume = {106},
    url = {https://link.aps.org/doi/10.1103/PhysRevD.106.126010},
    doi = {10.1103/PhysRevD.106.126010},
    number = {12},
    urldate = {2025-10-13},
    journal = {Physical Review D},
    author = {Hashimoto, Koji and Murata, Keiju and Tanahashi, Norihiro and Watanabe, Ryota},
    month = dec,
    year = {2022},
    pages = {126010},
}

@article{mertens_solvable_2023,
    title = {Solvable {Models} of {Quantum} {Black} {Holes}: {A} {Review} on {Jackiw}-{Teitelboim} {Gravity}},
    shorttitle = {Solvable {Models} of {Quantum} {Black} {Holes}},
    url = {http://arxiv.org/abs/2210.10846},
    urldate = {2023-06-22},
    archivePrefix = {arXiv},
    arxivID = {2210.10846},
    eprint = {2210.10846},
    author = {Mertens, Thomas G. and Turiaci, Gustavo J.},
    month = jun,
    year = {2023},
    note = {arXiv:2210.10846},
}

@article{martin_theory_1959,
    title = {Theory of {Many}-{Particle} {Systems}. {I}},
    volume = {115},
    url = {https://link.aps.org/doi/10.1103/PhysRev.115.1342},
    doi = {10.1103/PhysRev.115.1342},
    number = {6},
    urldate = {2025-10-13},
    journal = {Physical Review},
    author = {Martin, Paul C. and Schwinger, Julian},
    month = sep,
    year = {1959},
    pages = {1342--1373},
}

@book{horing_quantum_2017,
    title = {Quantum {Statistical} {Field} {Theory}: {An} {Introduction} to {Schwinger}'s {Variational} {Method} with {Green}'s {Function} {Nanoapplications}, {Graphene} and {Superconductivity}},
    isbn = {9780198791942},
    address ={Oxford, United Kingdom},
    shorttitle = {Quantum {Statistical} {Field} {Theory}},
    url = {https://doi.org/10.1093/oso/9780198791942.001.0001},
    urldate = {2025-10-13},
    publisher = {Oxford University Press},
    author = {Horing, Norman J. Morgenstern},
    month = jul,
    year = {2017},
    doi = {10.1093/oso/9780198791942.001.0001},
}

@article{seligman_scale-invariant_1985,
    title = {Scale-invariant {Lyapunov} exponents for classical hamiltonian systems},
    volume = {110},
    issn = {0375-9601},
    url = {https://www.sciencedirect.com/science/article/pii/0375960185900866},
    doi = {10.1016/0375-9601(85)90086-6},
    number = {5},
    urldate = {2025-10-13},
    journal = {Physics Letters A},
    author = {Seligman, T. H. and Verbaarschot, J. J. M. and Zirnbauer, M. R.},
    month = jul,
    year = {1985},
    pages = {231--234},
}

@book{KobayashiNomizu1969,
  author    = {Kobayashi, Shoshichi and Nomizu, Katsumi},
  title     = {Foundations of Differential Geometry, Volume II},
  publisher = {Interscience Publishers},
  address   = {New York},
  year      = {1969},
  isbn      = {0-471-09596-9},
}

@book{gutzwiller_chaos_1990,
    address = {New York, NY},
    series = {Interdisciplinary {Applied} {Mathematics}},
    title = {Chaos in {Classical} and {Quantum} {Mechanics}},
    volume = {1},
    copyright = {http://www.springer.com/tdm},
    isbn = {9781461269700 9781461209836},
    url = {http://link.springer.com/10.1007/978-1-4612-0983-6},
    urldate = {2025-10-13},
    publisher = {Springer},
    author = {Gutzwiller, Martin C.},
    editor = {John, F. and Kadanoff, L. and Marsden, J. E. and Sirovich, L. and Wiggins, S.},
    year = {1990},
    doi = {10.1007/978-1-4612-0983-6},
}

@book{sethna_statistical_2021,
    title = {Statistical {Mechanics}: {Entropy}, {Order} {Parameters}, and {Complexity}},
    isbn = {9780198865254},
    shorttitle = {Statistical {Mechanics}},
    address = {Oxford, United Kingdom},
    language = {en},
    publisher = {Oxford University Press},
    author = {Sethna, James P.},
    year = {2021},
}

@article{tsuji_bound_2018,
    title = {Bound on the exponential growth rate of out-of-time-ordered correlators},
    volume = {98},
    url = {https://link.aps.org/doi/10.1103/PhysRevE.98.012216},
    doi = {10.1103/PhysRevE.98.012216},
    number = {1},
    urldate = {2025-10-14},
    journal = {Physical Review E},
    author = {Tsuji, Naoto and Shitara, Tomohiro and Ueda, Masahito},
    month = jul,
    year = {2018},
    pages = {012216},
}

@article{Balian1970a,
    title = {Distribution of eigenfrequencies for the wave equation in a finite domain},
    volume = {60},
    issn = {00034916},
    url = {https://linkinghub.elsevier.com/retrieve/pii/0003491670904975},
    doi = {10.1016/0003-4916(70)90497-5},
    number = {2},
    urldate = {2022-02-03},
    journal = {Annals of Physics},
    author = {Balian, R. and Bloch, C.},
    month = oct,
    year = {1970},
    note = {Publisher: Academic Press},
    pages = {401--447},
}

@article{Balian1971,
    title = {Distribution of eigenfrequencies for the wave equation in a finite domain. {II}. {Electromagnetic} field. {Riemannian} spaces},
    volume = {64},
    issn = {00034916},
    url = {https://linkinghub.elsevier.com/retrieve/pii/0003491671902867},
    doi = {10.1016/0003-4916(71)90286-7},
    number = {1},
    urldate = {2022-02-03},
    journal = {Annals of Physics},
    author = {Balian, R. and Bloch, C.},
    month = may,
    year = {1971},
    note = {Publisher: Academic Press},
    pages = {271--307},
}

@article{Balian1972,
    title = {Distribution of eigenfrequencies for the wave equation in a finite domain: {III}. {Eigenfrequency} density oscillations},
    volume = {69},
    issn = {00034916},
    url = {https://linkinghub.elsevier.com/retrieve/pii/0003491672900061},
    doi = {10.1016/0003-4916(72)90006-1},
    number = {1},
    urldate = {2022-02-03},
    journal = {Annals of Physics},
    author = {Balian, R. and Bloch, C.},
    month = jan,
    year = {1972},
    note = {Publisher: Academic Press},
    pages = {76--160},
}

@phdthesis{steinhuber,
    author = {{Mathias Steinhuber}},
    title = {{Signatures of Instability in Bosonic Many-Body Systems}},
    school = {{University of Regensburg}},
    year = "2025"
}

@book{arnold_mathematical_1997,
    title = {Mathematical {Methods} of {Classical} {Mechanics}},
    isbn = {9780387968902},
    language = {en},
    publisher = {Springer Science \& Business Media},
    author = {Arnol'd, V. I.},
    month = sep,
    year = {1997},
}

@article{rammensee_many-body_2018,
    title = {Many-{Body} {Quantum} {Interference} and the {Saturation} of {Out}-of-{Time}-{Order} {Correlators}},
    volume = {121},
    url = {https://link.aps.org/doi/10.1103/PhysRevLett.121.124101},
    doi = {10.1103/PhysRevLett.121.124101},
    number = {12},
    urldate = {2025-11-20},
    journal = {Physical Review Letters},
    author = {Rammensee, Josef and Urbina, Juan Diego and Richter, Klaus},
    month = sep,
    year = {2018},
    pages = {124101},
}

@article{ehrenfest_bemerkung_1927,
    title = {Bemerkung \"{u}ber die angen\"{a}herte {G\"{u}ltigkeit} der klassischen {Mechanik} innerhalb der {Quantenmechanik}},
    volume = {45},
    issn = {0044-3328},
    url = {https://doi.org/10.1007/BF01329203},
    doi = {10.1007/BF01329203},
    language = {de},
    number = {7},
    urldate = {2025-11-20},
    journal = {Zeitschrift f\"{u}r Physik},
    author = {Ehrenfest, P.},
    month = jul,
    year = {1927},
    pages = {455--457},
}

@article{berman_condition_1978,
    title = {Condition of stochasticity in quantum nonlinear systems},
    volume = {91},
    issn = {0378-4371},
    url = {https://www.sciencedirect.com/science/article/pii/0378437178901905},
    doi = {10.1016/0378-4371(78)90190-5},
    number = {3},
    urldate = {2025-11-20},
    journal = {Physica A: Statistical Mechanics and its Applications},
    author = {Berman, G. P. and Zaslavsky, G. M.},
    month = may,
    year = {1978},
    pages = {450--460},
}

@book{minar_periodic_2024,
    title = {Periodic {Orbits} {Sums} in {Low}-dimensional {Quantum} {Systems}: {Convergence} {Conditions} and the {Maldacena}-{Shenker}-{Stanford} {Bound} on {Chaos}},
    shorttitle = {Periodic {Orbits} {Sums} in {Low}-dimensional {Quantum} {Systems}},
    language = {en},
    publisher = {University of Regensburg},
    author = {Min\'{a}r, Jakob Jan},
    year = {2024},
    note = {Google-Books-ID: chyq0AEACAAJ},
}

@article{richter_semiclassical_2022,
    title = {Semiclassical roots of universality in many-body quantum chaos},
    volume = {55},
    issn = {1751-8121},
    url = {https://dx.doi.org/10.1088/1751-8121/ac9e4e},
    doi = {10.1088/1751-8121/ac9e4e},
    language = {en},
    number = {45},
    urldate = {2023-03-28},
    journal = {Journal of Physics A: Mathematical and Theoretical},
    author = {Richter, Klaus and Urbina, Juan Diego and Tomsovic, Steven},
    month = nov,
    year = {2022},
    note = {Publisher: IOP Publishing},
    pages = {453001},
}

@article{Balazs1986,
  author       = {Balazs, N. L. and Voros, A.},
  title        = {Chaos on the pseudosphere},
  journal      = {Physics Reports},
  volume       = {143},
  number       = {3},
  pages        = {109--240},
  year         = {1986},
  doi          = {10.1016/0370-1573(86)90159-6},
}

@article{shenker_stringy_2015,
    title = {Stringy effects in scrambling},
    volume = {2015},
    issn = {1029-8479},
    url = {https://doi.org/10.1007/JHEP05(2015)132},
    doi = {10.1007/JHEP05(2015)132},
    language = {en},
    number = {5},
    urldate = {2025-12-09},
    journal = {Journal of High Energy Physics},
    author = {Shenker, Stephen H. and Stanford, Douglas},
    month = may,
    year = {2015},
    pages = {132},
}

@article{qi_quantum_2019,
    title = {Quantum correction to chaos in {Schwarzian} theory},
    volume = {2019},
    issn = {1029-8479},
    url = {https://doi.org/10.1007/JHEP11(2019)035},
    doi = {10.1007/JHEP11(2019)035},
    language = {en},
    number = {11},
    urldate = {2025-12-09},
    journal = {Journal of High Energy Physics},
    author = {Qi, Yong-Hui and Sin, Sang-Jin and Yoon, Junggi},
    month = nov,
    year = {2019},
    pages = {35},
}

@article{choi_effective_2023,
    title = {Effective description of sub-maximal chaos: stringy effects for {SYK} scrambling},
    volume = {2023},
    issn = {1029-8479},
    shorttitle = {Effective description of sub-maximal chaos},
    url = {https://doi.org/10.1007/JHEP03(2023)142},
    doi = {10.1007/JHEP03(2023)142},
    language = {en},
    number = {3},
    urldate = {2025-12-09},
    journal = {Journal of High Energy Physics},
    author = {Choi, Changha and Haehl, Felix M. and Mezei, Márk and S\'{a}rosi, Gábor},
    month = mar,
    year = {2023},
    pages = {142},
}

@article{butler_rigidity_2015,
    title = {Rigidity of equality of {Lyapunov} exponents for geodesic flows},
    url = {http://arxiv.org/abs/1501.05997},
    doi = {10.48550/arXiv.1501.05997},
    urldate = {2025-11-30},
    archivePrefix = {arXiv},
    arxivID = {1501.05997},
    eprint = {1501.05997},
    author = {Butler, Clark},
    month = oct,
    year = {2015},
    note = {arXiv:1501.05997},
}

@article{pappalardi_low_2022,
    title = {Low temperature quantum bounds on simple models},
    volume = {13},
    issn = {2542-4653},
    url = {https://scipost.org/10.21468/SciPostPhys.13.1.006},
    doi = {10.21468/SciPostPhys.13.1.006},
    language = {en},
    number = {1},
    urldate = {2025-12-21},
    journal = {SciPost Physics},
    author = {Pappalardi, Silvia and Kurchan, Jorge},
    month = jul,
    year = {2022},
    pages = {006},
}

@article{pappalardi_quantum_2023,
    title = {Quantum {Bounds} on the {Generalized} {Lyapunov} {Exponents}},
    volume = {25},
    copyright = {http://creativecommons.org/licenses/by/3.0/},
    issn = {1099-4300},
    url = {https://www.mdpi.com/1099-4300/25/2/246},
    doi = {10.3390/e25020246},
    language = {en},
    number = {2},
    urldate = {2025-12-21},
    journal = {Entropy},
    author = {Pappalardi, Silvia and Kurchan, Jorge},
    month = feb,
    year = {2023},
    pages = {246},
}

\appendix
\section{Accessing the classical regime}\label{app:regime}
We consider the product
\begin{equation}
    \prod_{k=0}^{\frac{f-3}{2}}\left(\frac{2mL^2}{\hbar^2}E+k^2\right)\equiv\prod_{k=0}^M(x+k^2)=x^M\prod_{k=0}^M\left(1+\frac{k^2}{x}\right).
\end{equation}
For this product to be dominated by the highest-degree monomial, we need
\begin{equation}
     \prod_{k=0}^M\left(1+\frac{k^2}{x}\right)\approx1,
\end{equation}
which is true when
\begin{equation}
    \abs{\prod_{k=0}^M\left(1+\frac{k^2}{x}\right)-1}\leq e^{\sum_{k=0}^M\frac{k^2}{x}}-1\approx0,
\end{equation}
i.e.
\begin{equation}
    \sum_{k=0}^M\frac{k^2}{x}\ll1.
\end{equation}
This sum can be evaluated exactly as
\begin{equation}
    \sum_{k=0}^Mk^2=\frac{M(M+1)(2M+1)}{6}=\frac{(f-3)(f-2)(f-1)}{24}\approx\frac{f^3}{24}
\end{equation}
for large $f$. Hence, the condition to access the classical regime in our system is
\begin{equation}
    \frac{f^3}{24}\ll\frac{2mL^2}{\hbar^2}E,
\end{equation}
and if we assume that in the classical regime, the energy is given roughly by the thermal energy,
\begin{equation}
    E\approx\frac{f}{\beta},
\end{equation}
we obtain the classicality condition reported in the main text,
\begin{equation}
    \frac{f^2}{24}\ll\frac{2mL^2}{\hbar^2\beta}=\frac{4\pi L^2}{\lambda^2_\mathrm{th}}.
\end{equation}
\section{Weyl symbol of the Hamiltonian}\label{app:weyl-symbol}
To find the Weyl symbol for our Hamiltonian \eqref{eq:hadamard-gutzwiller-hamiltonian}, consider
\begin{align}
    \frac{2m}{\hbar^2}H&=-\Delta=-g^{-1/2}\partial_ig^{1/2}g^{ij}\partial_j\nonumber\\
    &=-g^{-1/2}(\partial_ig^{1/2}g^{ij})\partial_j-g^{ij}\partial_i\partial_j\nonumber\\
    &=-\partial^2-g^{-1/2}(\partial_ig^{1/2})g^{ij}\partial_j-(\partial_ig^{ij})\partial_j\nonumber\\
    &=-\partial^2-(\partial_ig^{ij})\partial_j-(\partial_i\log\sqrt{g})g^{ij}\partial_j\nonumber\\
    &=-\partial^2-(\partial_ig^{ij})\partial_j-\frac{1}{2}g^{ij}\bigg[(\partial_i\log\sqrt{g})\partial_j+(\partial_j\log\sqrt{g})\partial_i\bigg].\label{eq:laplace-rewritten}
\end{align}
Compare this to DeWitt's ordering \cite{dewitt_point_1952,dewitt_dynamical_1957,gneiting_quantum_2013},
\begin{equation}
    H_{\text{DeWitt}}=\frac{1}{2m}P_ig^{ij}(X)P_j+\hbar^2Q(X),
\end{equation}
with the momentum operator
\begin{equation}
    P_i=\frac{\hbar}{i}\left(\partial_i+\frac{1}{2}\Gamma^j_{ji}(x)\right)
\end{equation}
and the so-called quantum potential
\begin{equation}
    Q(x)=\frac{1}{4m}g^{ij}\left[\partial_j\Gamma^k_{ki}-\Gamma^k_{ij}\Gamma^l_{lk}-\frac{1}{2}\Gamma^k_{ki}\Gamma^l_{lj}\right].
\end{equation}
$\Gamma^k_{ij}$ are the standard Christoffel symbols. First, note that using
\begin{equation}
    \Gamma_i=\Gamma^j_{ji}=\partial_i\log\sqrt{g},\label{eq:christoffels}
\end{equation}
the quantum potential can be rewritten as \cite{marinov_path_1980}
\begin{equation}
    Q(x)=\frac{1}{4m}\partial_i(g^{ij}\Gamma_j)+\frac{1}{8m}g^{ij}\Gamma_i\Gamma_j.
\end{equation}
Furthermore, the Hamiltonian
\begin{align}
    \frac{2m}{\hbar^2}H_{\text{DeWitt}}&=2mQ(x)-\left(\partial_i+\frac{1}{2}\Gamma_i(x)\right)g^{ij}(x)\left(\partial_j+\frac{1}{2}\Gamma_j(x)\right)\nonumber\\
    &=2mQ-\frac{1}{2}(\partial_ig^{ij})\Gamma_j-\frac{1}{4}\Gamma_ig^{ij}\Gamma_j-\frac{1}{2}(\partial_i\Gamma_j)g^{ij}\nonumber\\
    &\hphantom{=}-(\partial_ig^{ij})\partial_j-\partial^2-\frac{1}{2}g^{ij}\Gamma_j\partial_i-\frac{1}{2}\Gamma_ig^{ij}\partial_j\nonumber\\
    &=-(\partial_ig^{ij})\partial_j-\partial^2-\frac{1}{2}g^{ij}\left(\Gamma_j\partial_i+\Gamma_i\partial_j\right)\label{eq:dewitt-rewritten}
\end{align}
agrees with \cref{eq:laplace-rewritten}. Fortuitously, the Weyl symbol for the DeWitt Hamiltonian is reported in \cite{gneiting_quantum_2013} as
\begin{equation}
    W_{H_{\text{DeWitt}}}(x,p)=W_H(x,p)=\frac{1}{2m}p_ig^{ij}(x)p_j+\hbar^2Q(x)+\frac{\hbar^2}{8m}\partial_i\partial_jg^{ij}(x).
\end{equation}
After changing into Riemannian normal coordinates, i.e. coordinates such that at a point $q$, $g_{ij}(q)=\delta_{ij}$, $\partial_kg_{ij}(q)=0$, $\Gamma^k_{ij}(q)=0$, but $\partial_l\Gamma^k_{ij}(q)\neq0$, a somewhat tedious calculation shows that the $\hbar^2$ correction to the Weyl symbol of the Hamiltonian is simply given by the Ricci scalar of the manifold,
\begin{equation}
    h_2=\frac{R}{12m}.
\end{equation}
\end{document}